\documentclass[a4paper,11pt]{article}
\pdfoutput=1 

\usepackage{jheppub} 

\usepackage{verbatim} 
\usepackage{url}

\def\lr{\left( }
\def\rr{\right) }
\def\le{\left[ }
\def\re{\right] }

\usepackage{soul}

\newcommand{\beq}{\begin{equation}}
\newcommand{\eeq}{\end{equation}}
\newcommand{\bea}{\begin{eqnarray}}
\newcommand{\eea}{\end{eqnarray}}

\newcommand{\fcse}{f_\text{CSE}}

\newcommand{\pTj}{p_\text{T}^\text{jet2}}
\newcommand{\Deta}{\Delta\eta_\text{jets}}
\newcommand{\Dphi}{\Delta\phi_\text{jets}}

\newcommand{\intc}[1]{{\int\frac{d#1}{2i\pi}}}

\preprint{MS-TP-22-11}

\title{Jets separated by a large pseudorapidity gap at the Tevatron and at the LHC}

\author[a]{C.~Baldenegro,}
\author[b]{P.~Gonz{\'a}lez Dur{\'a}n,}
\author[b]{M.~Klasen,}
\author[c]{C.~Royon}
\author[b]{and J.~Salomon}

\affiliation[a]{\'{E}cole Polytechnique, Laboratoire Leprince-Ringuet, Av. Chasles, 91120 Palaiseau, France}
\affiliation[b]{Westfälische Wilhelms-Universität Münster, Institut f{\"u}r Theoretische Physik, Wilhelm-Klemm-Str. 9, 48149 M{\"u}nster, Germany}
\affiliation[c]{The University of Kansas, Department of Physics and Astronomy, 1082 Malott Hall, 1251 Wescoe Hall Drive, Lawrence, KS~66045, U.S.A.}

\emailAdd{c.baldenegro@cern.ch}
\emailAdd{paegodu@gmail.com}
\emailAdd{michael.klasen@uni-muenster.de}
\emailAdd{christophe.royon@ku.edu}
\emailAdd{jens.salomon@uni-muenster.de}

\abstract{We present a phenomenological analysis of events with two high transverse momentum ($p_T$) jets separated by a large (pseudo-)rapidity interval void of particle activity, also known as jet-gap-jet events. In the limit  where the collision energy $\sqrt{s}$ is much larger than any other momentum scale, the jet-gap-jet process is described in terms of perturbative pomeron exchange between partons within the Balitsky--Fadin--Kuraev--Lipatov (BFKL) limit of perturbative quantum chromodynamics (QCD). The BFKL pomeron exchange amplitudes, with resummation at the next-to-leading logarithmic approximation, have been embedded in the PYTHIA8 Monte Carlo event generator. Standard QCD dijet events are simulated at next-to-leading order in $\alpha_s$ matched to parton showers with POWHEG+PYTHIA8. We compare our calculations to measurements by the CDF, D0, and CMS experiments at center-of-mass energies of 1.8, 7 and 13 TeV. The impact of the theoretical scales, the parton densities, final- and initial-state radiation effects, multiple parton interactions, and $p_T$ thresholds and multiplicities of the particles in the rapidity gap on the jet-gap-jet signature is studied in detail. With a strict gap definition (no particle allowed in the gap), the shapes of most distributions are well described except for the CMS azimuthal-angle distribution at 13 TeV. 
The survival probability is surprisingly well modelled by multiparton interactions in PYTHIA8. Without multiparton interactions, theoretical predictions based on two-channel eikonal models agree qualitatively with fits to the experimental data.}

\begin{document} 
\maketitle
\flushbottom

\section{Introduction}
\label{sec:1}

The production of high transverse momentum ($p_T$) jets, i.e.\ collimated sprays of particles, in high-energy ($\sqrt{s}$) hadron-hadron collisions is well-described by quantum chromodynamics (QCD), the quantum field theory of the strong interaction. In perturbative QCD (pQCD), the parton-level cross section can be calculated order-by-order in a series expansion in the strong coupling $\alpha_s \ll 1$. The parton-level cross section is then convolved with universal parton distribution functions (PDFs) to obtain the absolute value of the cross section in hadron-hadron collisions. These calculations at leading order (LO), next-to-leading order (NLO) in $\alpha_s$ and beyond can furthermore be supplemented with collinear parton emissions in the parton shower (PS) algorithms embedded in Monte Carlo (MC) event generators, which results in the production of multiple partons originating from the elementary parton splitting functions of QCD. At some point, there is a transition of degrees of freedom from partons to hadrons, known as the hadronization process, which is taken into account with QCD-inspired models that are tuned to collider data. For the phase-space regions explored by the Tevatron and LHC experiments, this approach works remarkably well and has served as an excellent testing and validation ground of QCD. Nevertheless, there are good reasons to expect that the fixed-order pQCD approach used for the parton-level cross sections should break down in special multijet configurations \cite{Ellis:1996mzs,Forshaw:1997dc}.

\begin{figure}
\centering
\includegraphics[width=.4\textwidth]{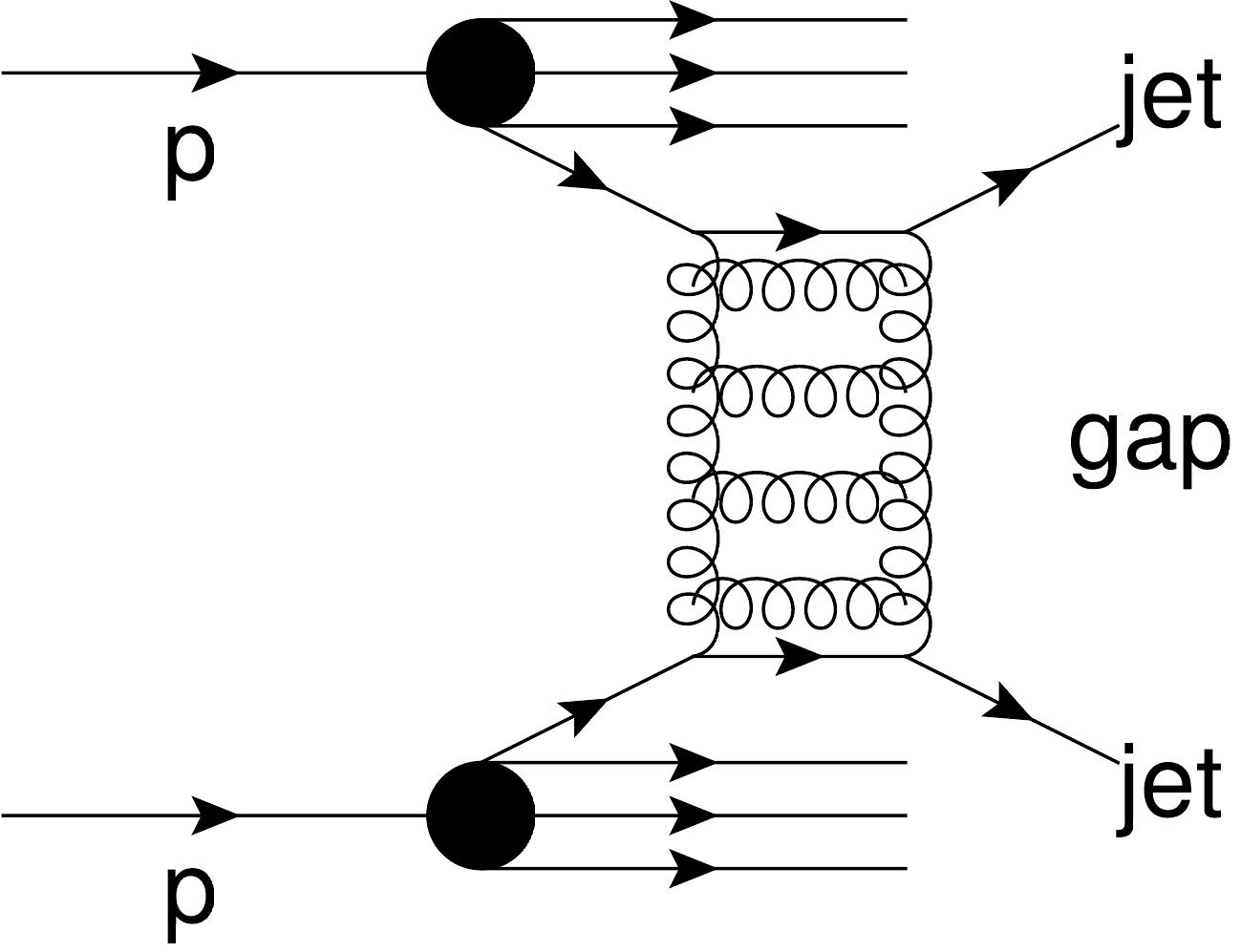}
\hspace{1cm}
\includegraphics[width=.45\textwidth]{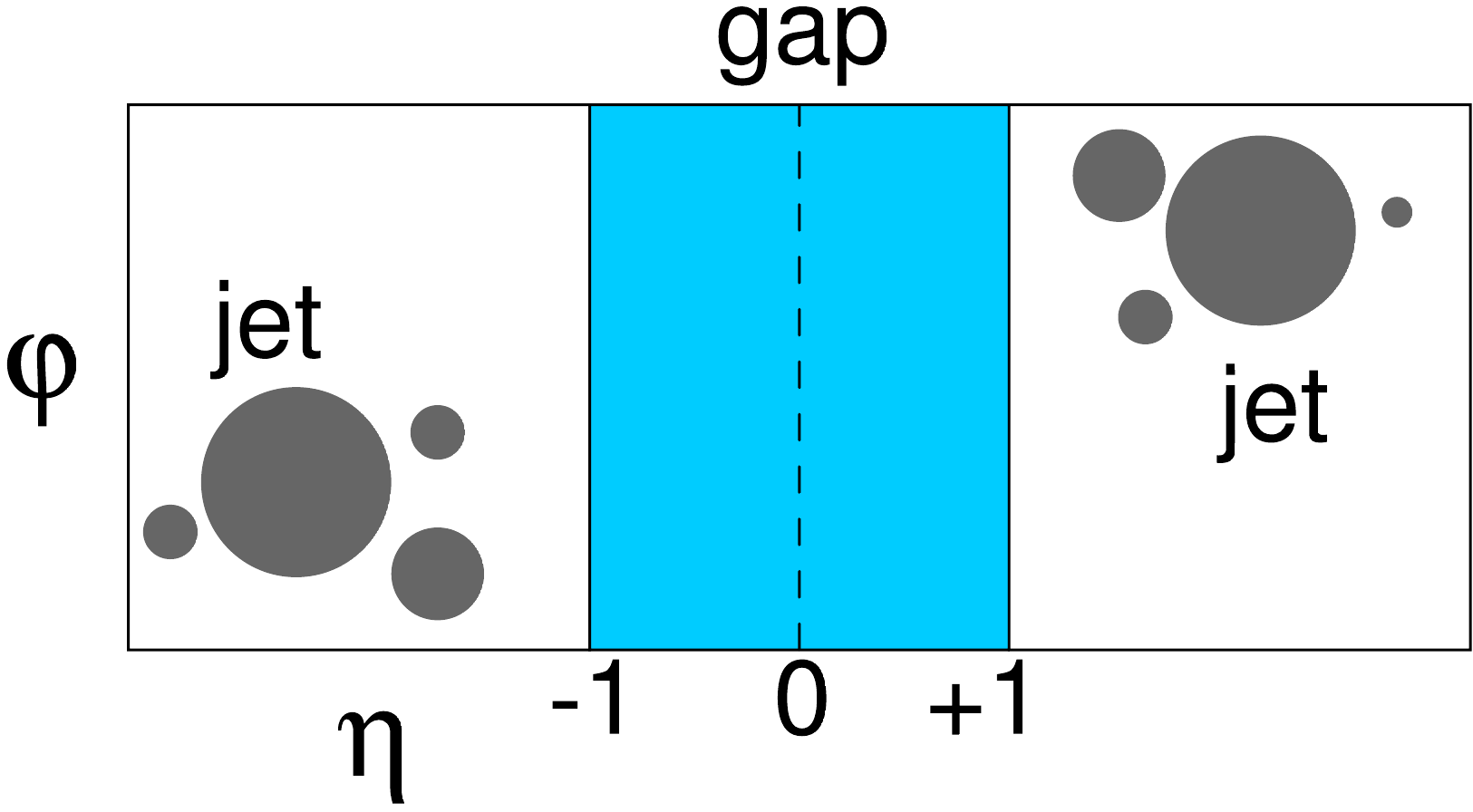}
\caption{Left: Schematic diagram of the production of two jets separated by a large rapidity gap in a hadron-hadron collision via color-singlet exchange between quarks and gluons. Right: Signature of jet-gap-jet events in the (pseudo-)rapidity--azimuthal angle ($\eta$--$\phi$) plane. The blue shaded area represents the interval in pseudorapidity void of particles. Both diagrams are extracted from Ref.~\cite{Sirunyan:2021oxl}.}
\label{fig:01}
\end{figure}

In this paper, we focus on events with two high-$p_T$ jets separated by a large interval in (pseudo-)rapidity $\eta$ void of any particle activity, known as jet-gap-jet or Mueller--Tang jet events (cf.\ Fig.\ \ref{fig:01}) \cite{Mueller:1992pe}. This topology is consistent with an underlying $t$-channel color-singlet exchange between partons, which is generated at leading order in pQCD by a two-gluon exchange. This is in contrast to color-octet (or higher multiplet) exchanges between partons, where the net color exchange between partons leads to the production of several soft hadrons between the high-$p_T$ jets. The latter dominates the inclusive dijet cross section. In the regime where the two jets are largely separated in rapidity, higher-order corrections to the aforementioned color-singlet two-gluon exchange need to be taken into account. This is due to diagrams with additional virtual gluon exchanges, which become very important in the high-energy limit of QCD. These diagrams contribute to the scattering amplitude with terms proportional to $\alpha^n_s \log^n (\hat{s}) \lesssim \mathcal{O}(1)$ in the perturbative expansion, which need to be resummed. Here, $\hat{s}$ is the center-of-mass energy of the partonic process. Resummation of these leading logarithmic (LL), but also next-to-LL (NLL) terms to all orders is achieved via the Balitsky--Fadin--Kuraev--Lipatov (BFKL) evolution equation of pQCD \cite{Kuraev:1977fs, Balitsky:1978ic}. The resulting color-singlet exchange is known as perturbative pomeron exchange, i.e.\ a $t$-channel ladder exchange of reggeized gluons. Thus, the jet-gap-jet process is a promising gateway to identify the onset of BFKL dynamics in the data. At the same time, in events with large rapidity gaps the contributions of other higher-order corrections in QCD, such as the Dokshitzer--Gribov--Lipatov--Altarelli--Parisi (DGLAP) dynamics, are intrinsically suppressed~\cite{dglap1,dglap2,dglap3}. The reason is that, by requiring a large rapidity gap between the jets, one is effectively suppressing any higher-order correction that yields radiation between the jets by means of a Sudakov form factor. Thus, the process is clean from the point of view of the short-distance mechanism responsible for the production of the jet-gap-jet signature.

Unfortunately, there is no unique way of defining the rapidity gap between the jets. Experimentally, it may be defined as the absence of particles above a non-zero $p_T$ threshold (typically 200--300 MeV) \cite{Abe:1998ip,Abe:1997ie,Abbott:1998jb,Abbott:1999ai,Sirunyan:2017rdp,Sirunyan:2021oxl}, which is close to the detector noise threshold or to the regime where particle reconstruction becomes difficult. The residual color-octet dijet background is subtracted in these measurements. On the other hand, from a theoretical perspective one would prefer to bring this $p_T$ threshold in the experiment as low as possible at hadron-level. Therefore, it is worth examining the role of the $p_T$ threshold used in the rapidity gap definition in the context of underpinning the underlying hard scattering mechanism. In this paper, we compare two different gap definitions to isolate the CSE dijet events: a ``strict gap'' definition (no particles, regardless of their $p_T$) and a definition that is as close as possible to the one adopted by the experiments, which we dub ``experimental gap.'' We compare the results of these two approaches with the bare CSE dijet cross section prediction without a rapidity gap requirement to have a common reference point.

The experimental observable that is usually extracted is the fraction of dijet events produced by a color-singlet exchange (CSE),
\begin{equation}
    f_\text{CSE} \equiv \sigma^\text{CSE}/ \sigma^\text{inc},
\end{equation}
where $\sigma^\text{CSE}$ and $\sigma^\text{inc}$ are the cross sections for color-singlet exchange dijet events and inclusive dijet events, respectively. The latter includes the contributions of color-singlet exchange as well, although it is expected to be largely dominated by color-octet exchange. Such a ratio-based quantity was suggested in the original paper by Mueller and Tang~\cite{Mueller:1992pe}. The ratio defined this way leads to the cancellation of various correlated uncertainties related to jet energy scale corrections, parton shower, hadronization, and parton distribution functions of the CSE and inclusive dijet events. Such an observable is the focus of this paper.

Phenomenological analyses of Mueller--Tang jets have been presented in Refs.~\cite{Cox:1999dw, Motyka:2001zh, Chevallier:2009cu, Kepka:2010hu, Babiarz:2017jxc}. The analysis in Ref.~\cite{Chevallier:2009cu} considered the phenomenological description of the jet-gap-jet process at parton level, whereas the following analysis in Ref.~\cite{Kepka:2010hu} took into account parton showering and hadronization effects through an implementation of the process in the HERWIG6 MC event generator~\cite{Corcella:2000bw}. We refer to the latter study at hadron-level as the Royon--Marquet--Kepka (RMK) calculation hereafter. The use of the BFKL gluon Green's function with the NLL corrections and summed over conformal spins was first introduced in Refs.\ \cite{Chevallier:2009cu,Kepka:2010hu} to make predictions for the Mueller--Tang jet process. The impact factor, which describes how the perturbative pomeron exchange couples to the initial-state partons from the colliding protons, was calculated at leading order (LO). The gap survival probability $\mathcal{S}_\mathrm{prob}$ was implemented with a multiplicative, static constant, which allowed to fit the normalization of the data. For inclusive dijet production, next-to-leading order (NLO) QCD calculations at parton-level were used to reweight the MC events \cite{Nagy:2001xb}.

Another set of predictions was presented by the group of Ekstedt--Enberg--Ingelman--Motyka (EEIM) ~\cite{csp, cspLHC}. Here, the dominant NLL corrections to the BFKL gluon Green's function were implemented in their calculation by restricting the momentum of the real gluon emission in the BFKL eigenvalue. The effect from higher conformal spins was implemented with an approximation that accounts for the modification of the partonic cross section at medium pseudorapidity differences between the jets. The LO impact factor for the Mueller--Tang process was used in these calculations as well. In the calculations by the EEIM group, the survival probability $\mathcal{S}_\mathrm{prob}$ was dynamically calculated with a simulation of the underlying event activity and the soft color interaction model. The calculations were embedded as a subroutine in the PYTHIA6 MC event generator \cite{Sjostrand:2006za}. For the inclusive dijet cross section, LO QCD calculations with parton showers in PYTHIA6 were employed.

Previous phenomenological predictions \cite{Cox:1999dw,csp,Kepka:2010hu} were compared to the available CDF and D0 measurements of jet-gap-jet events in proton-antiproton collisions at $\sqrt{s} = 0.63$ and $1.8$ TeV \cite{Abe:1998ip,Abe:1997ie,Abbott:1998jb,Abbott:1999ai}. Theory predictions are in fair agreement with experimental measurements 
within the experimental uncertainties. However, the measurements by the Tevatron experiments were somewhat limited statistically and did not allow for a conclusive discrimination between the BFKL-only predictions and the DGLAP-based pQCD dynamics. At the same time, the Tevatron data did not cover a wide region of phase-space that could be used to test the BFKL dynamics, particularly the region of large $\Delta\eta_\text{jets} \equiv |\eta_\mathrm{1}-\eta_\mathrm{2}|$ between to the two highest $p_\mathrm{T}$ jets in the event. Recently, the CMS collaboration at the LHC has presented measurements of the jet-gap-jet process in proton-proton ($pp$) collisions at $\sqrt{s} =7$ TeV \cite{Sirunyan:2017rdp} and 13 TeV \cite{Sirunyan:2021oxl}, which has complemented the phase-space region previously probed by the Tevatron experiments with increased statistical accuracy. The aforementioned theoretical calculations have not yet been compared in detail to the LHC data, and they did not also  account for up-to-date parametrizations of the underlying event activity, initial- and final-state radiation effects, as well as fragmentation models that have now been tuned to the LHC Run-1 and Run-2 data at larger $\sqrt{s}$. For this work, we therefore implemented the jet-gap-jet process at NLL accuracy in PYTHIA8 \cite{Sjostrand:2014zea}, a modern MC event generator that has been tuned to LHC data, and simulated inclusive dijet events at NLO matched to parton showers with POWHEG+PYTHIA8 \cite{Nason:2004rx,Frixione:2007vw,Alioli:2010xa,Alioli:2010xd}. We compare our predictions to both the Tevatron and the new LHC measurements and perform a systematic study of various relevant effects that enter the jet-gap-jet prediction. The PYTHIA8 subroutine for Mueller--Tang jets is publicly available~\cite{githublink}.

The paper is structured as follows. The theoretical framework used for our NLL BFKL and NLO pQCD calculations as well as our implementation of the jet-gap-jet process in PYTHIA8 are described in Section~\ref{sec:2}. Section~\ref{sec:3} contains our main numerical results for jet-gap-jet events at the Tevatron and the LHC with detailed studies of the effects of the gap definition, parton showers and multi-parton interactions described therein and in Section~\ref{sec:4}. The resulting scale factors are compared with theoretical predictions in Sec.\ \ref{sec:5}. In Section~\ref{sec:6}, we present a summary and an outlook. The appendix contains technical details of the NLL BFKL implementation and numerical results for the inclusive dijet cross section for future reference.

\section{Theoretical framework}
\label{sec:2}

We now describe in detail our theoretical framework, i.e.\ our new implementation of the BFKL process in PYTHIA8, our calculations of NLO pQCD dijet production matched to parton showers with the POWHEG BOX \cite{Nason:2004rx,Frixione:2007vw,Alioli:2010xa,Alioli:2010xd}, the validation of our new setup against inclusive dijet measurements and the previous RMK results at the Tevatron, as well as our choices of underlying event tunes and PDFs.

\subsection{Mueller--Tang jet implementation at NLL+PS in PYTHIA8}
\label{sec:2.1}

In this paper, the cross section for Mueller--Tang jet-gap-jet events is calculated in a similar way as in the RMK approach \cite{Kepka:2010hu}. There, the differential CSE cross section was expressed in the form
\beq
\frac{d \sigma^{\rm CSE}}{dx_1 dx_2 dp_T^2} = \mathcal{S}_\mathrm{prob}f_{\rm eff}(x_1,p_T^2)f_{\rm eff}(x_2,p_T^2)
\frac{d \sigma^{gg\to gg}}{dp_T^2},
\label{jgj}
\eeq
where $\frac{d \sigma^{gg\to gg}}{dp_T^2}$ denotes the parton-level cross section for gluon-gluon scattering and where the functions $f_{\rm eff}(x,p_T^2)$ are the effective PDFs that account for the color factors associated to the CSE process in QCD, namely
\beq
f_{\rm eff}(x,\mu^2)=g(x,\mu^2)+\frac{C_F^2}{N_c^2} \sum_{\rm f} \lr q_f(x,\mu^2)+\bar{q_f}(x,\mu^2)\rr\ .
\label{pdfs}
\eeq
Here, $q_f(x,\mu^2)$ and $g(x,\mu^2)$ denote the quark and gluon PDFs as a function of the parton momentum fraction $x$ at a scale $\mu$, and $C_A = N_c = 3$ and $C_F = 4/3$. The sum runs over all possible quark flavors $f$. In the BFKL limit, the $2\to 2$ parton-level cross section is the same for quark-(anti)quark ($qq \to qq$), (anti)quark-gluon ($qg \to qg$), and gluon-gluon ($gg \to gg$) scattering up to a color factor, as expressed explicitly in Eqs.~\eqref{jgj} and \eqref{pdfs}.

The parton-level cross section for $gg\to gg$ scattering is given by
\beq
\frac{d \sigma^{gg\to gg}}{dp_T^2}=\frac{1}{16\pi} \left|A(\Delta y,p_T^2)\right|^2,
\eeq
where $A$ is the scattering amplitude and $\Delta y$ is the difference in rapidity between the two outgoing gluons. Applying the Mueller--Tang prescription at NLL leads to
\beq
A(\Delta y ,p_T^2)=\frac{N_c}{C_F} \frac{16 \pi\alpha_s^2}{ p_T^2}\sum_{p=-\infty}^\infty\!\intc{\gamma}\!
\frac{[p^2-(\gamma-1/2)^2]\exp\left\{\bar\alpha(p_T^2)\chi_{\rm eff}[2p,\gamma,\bar\alpha(p_T^2)] \Delta y \right\}}
{[(\gamma-1/2)^2-(p-1/2)^2][(\gamma-1/2)^2-(p+1/2)^2]},
\label{jgjnll}
\eeq
where the complex integral in $\gamma$ is calculated along the imaginary axis from $1/2\!-\!i\infty$ to $1/2\!+\!i\infty,$ and where we sum over even conformal spins \cite{Motyka:2001zh}, which are represented by $p$ under the sum. We use the effective coupling constant $\bar{\alpha} = \alpha_s N_c/\pi$. At NLL, $\alpha_s$ runs with the hard energy scale of the process (given by the transverse momentum of the jets in our calculations) following the renormalization group equation. We use the resummation scheme S4 \cite{Salam:1998tj} for our calculations. This allows us to extend the regularization procedure to non-zero conformal spins and obtain  $\chi_\mathrm{NLL}\lr p,\gamma,\omega\rr$ \cite{Salam:1998tj}. Then, the effective kernels  $\chi_{\rm eff}(p,\gamma,\bar\alpha)$ are obtained from the NLL kernel by solving the implicit equation that relates the effective kernels to the NLL ones as described in Ref.\ \cite{Kepka:2010hu}. Details on the parametrization are presented in App.\ \ref{app:a} of this paper.

Technically, we implemented three new hard $2\to 2$ processes ($gg \to gg$, $qg \to qg$ and $qq \to qq$) corresponding to $t$-channel perturbative pomeron exchange as ``semi-internal processes'' in PYTHIA8 \cite{Sjostrand:2014zea} in order to account for the correct PDFs and color factors in Eq.~(\ref{jgj}) as well as for the correct color flow topologies, making use of its internal phase space selection to sample externally provided cross-sections. These gluon~($g$) and quark/antiquark~($q$) subprocesses share the routine based on the functional parametrization of the differential cross section of the numerical BFKL calculations at NLL, as it was done in Ref.~\cite{Kepka:2010hu}. In contrast to the previous RMK study \cite{Kepka:2010hu}, where the angular-ordered parton shower of HERWIG6 was used that automatically ensures color coherence \cite{Corcella:2000bw}, initial- and final-state parton showers are calculated here with PYTHIA8, i.e.\ ordered in $p_T$, combined with a dipole-style phase space to ensure color coherence and interleaved with multi-parton interactions (MPI) \cite{Sjostrand:2004ef}. Hadronization is simulated with the Lund string fragmentation model \cite{LundString}.

The cross section in Eq.~\eqref{jgj} does not obey collinear factorization. This is due to possible secondary soft interactions between the colliding hadrons which can fill the rapidity gap between the jets. Therefore, in Eq.~\eqref{jgj}, the collinear factorization of the parton distributions $f_{\rm eff}$ is corrected with the gap survival probability $\mathcal{S}_\mathrm{prob}$. The simplest approach is to take $\mathcal{S}_\mathrm{prob}$ as an additional, multiplicative absolute normalization factor that is fitted to data and assume that it only depends on $\sqrt{s}$ as in standard diffractive processes. As the soft interactions happen on much longer time scales, it is assumed that the factor $\mathcal{S}_\mathrm{prob}$ can be factorized from the hard cross section. Alternatively, the survival probability can be modelled dynamically, i.e.\ with multiparton interactions. Striclty speaking, the survival probability is process-dependent and also depends on the definition of the rapidity gap, as described for example in Ref.~\cite{Khoze:2013dha}.

In the following, we distinguish two quantities related to the overall normalization of the cross section: the survival probability $\mathcal{S}_\mathrm{prob}$ (which in principle is calculable with nonperturbative QCD techniques) and an overall scale factor $\mathcal{S}$. The latter is used to normalize our calculations to better describe the data. $\mathcal{S}$ contains the effects in $\mathcal{S}_\mathrm{prob}$, but also additional factors related to higher-order corrections that are currenty missing in our calculations. For instance, the BFKL dijet cross section will change with the inclusion of NLO impact factors \cite{hentschinski1,hentschinski2}, thus requiring a different scale factor $\mathcal{S}$ to describe the data. On the other hand, the nonperturbative $\mathcal{S}_\mathrm{prob}$ should be independent of such corrections. Later in this paper, we discuss how the scale factor $\mathcal{S}$ depends on different observables and $\sqrt{s}$.

\subsection{Inclusive dijet production at NLO+PS in POWHEG+PYTHIA8}
\label{sec:2.2}

NLO QCD cross section predictions and simulations of inclusive dijet events are obtained with the dijet implementation in the POWHEG BOX \cite{Nason:2004rx,Frixione:2007vw,Alioli:2010xa,Alioli:2010xd} and matched to the parton showers of PYTHIA8 \cite{Sjostrand:2014zea}. The POWHEG BOX is a general computer framework for implementing NLO calculations in MC generators according to the Positive Weight Hardest Emission Generator (POWHEG) method. It allows to avoid double-counting of the fixed-order NLO corrections with the NLO corrections that are approximated by parton shower algorithms. Initial- and final-state parton showers are again calculated with PYTHIA8, i.e.\ ordered in $p_T$, combined with a dipole-style phase space to ensure color coherence and interleaved with multi-parton interactions (MPI) \cite{Sjostrand:2004ef}. Hadronization is again simulated with the Lund string fragmentation model \cite{LundString}. Details of the PYTHIA8 PS and MPI simulation have of course some importance for the inclusive dijet cross section (see App.\ \ref{app:b}), but their impact is much larger for the Mueller--Tang jet-gap-jet cross section.

The fraction of color-singlet exchange events is obtained from the ratio
\begin{equation}
f_\text{CSE} = \frac{\text{NLL BFKL in PYTHIA8}}{\text{NLO+PS with POWHEG+PYTHIA8}}
\end{equation}
at hadron-level after clustering all particles into jets. Experimentally, the denominator consists of a mixture of color-octet, or higher multiplets, and color-singlet exchange. However, in global PDF fit analyses based on inclusive jet cross section measurements, no such distinctions are made; all jets are counted inclusively, regardless of the underlying topology. Thus, theoretically the inclusive dijet cross section in the denominator includes implicitly the contributions from CSE events. In fact, incorrectly including the CSE contribution in the denominator in addition to the NLO QCD contribution significantly changes the predicted $f_\text{CSE}$ behavior. 

\subsection{Validation of the NLO+PS and BFKL implementations}
\label{sec:2.3}

As a first step, we validated our absolute NLO QCD dijet cross section obtained with POWHEG+PYTHIA8 by comparing our results for the invariant dijet mass ($M$) distribution $d^3\sigma/dM\,d\eta_1\,d\eta_2$ in $p\bar{p}$ collisions at $\sqrt{s}=1.8$ TeV with the measurements by the D0 collaboration at the Tevatron and the NLO QCD results obtained with JETRAD \cite{Giele:1993dj} as presented in Fig.\ 1 of Ref.\ \cite{D0:1998byg}. Using the same settings for the renormalization and factorization scales ($\mu_R=\mu_F=0.5\,p_T$) and similar PDFs (the set CTEQ6M \cite{cteq6l1} still available in Version 5 of LHAPDF \cite{Buckley:2014ana} closest to the older set CTEQ3M \cite{Lai:1994bb} used in Ref.\ \cite{D0:1998byg}), we obtained very good agreement in absolute size and shape of the cross section as well as for the scale uncertainty, quoted as 30\% in Ref.\ \cite{D0:1998byg}.

The inclusive dijet cross section discussed above was chosen to be similar in experimental conditions to the jet-gap-jet measurements by the D0 collaboration published shortly thereafter \cite{Abbott:1998jb}. The latter were given with absolute normalization of $f_{\rm CSE}$, in contrast to the CDF data \cite{Abe:1997ie}, which were normalized to be unity on average. Both data sets were compared to in the RMK analysis \cite{Kepka:2010hu}. From the absolute BFKL cross sections for D0 conditions in Ref.\ \cite{Chevallier:2009cu} and the ratios $f_{\rm CSE}$ in Refs.\ \cite{Chevallier:2009cu,Kepka:2010hu}, we were able to reconstruct and reproduce their inclusive NLO QCD distributions in $p_T$ and dijet pseudorapidity difference ($\Delta\eta_{\rm jets}$) in the low- and high-$p_T$ bins in shape. 
We therefore validated both shape and magnitude also against our own NLO QCD program \cite{Klasen:1996yk,Klasen:1996it,Klasen:1997br}.

\begin{figure}
\centering
\includegraphics[width=0.53\linewidth]{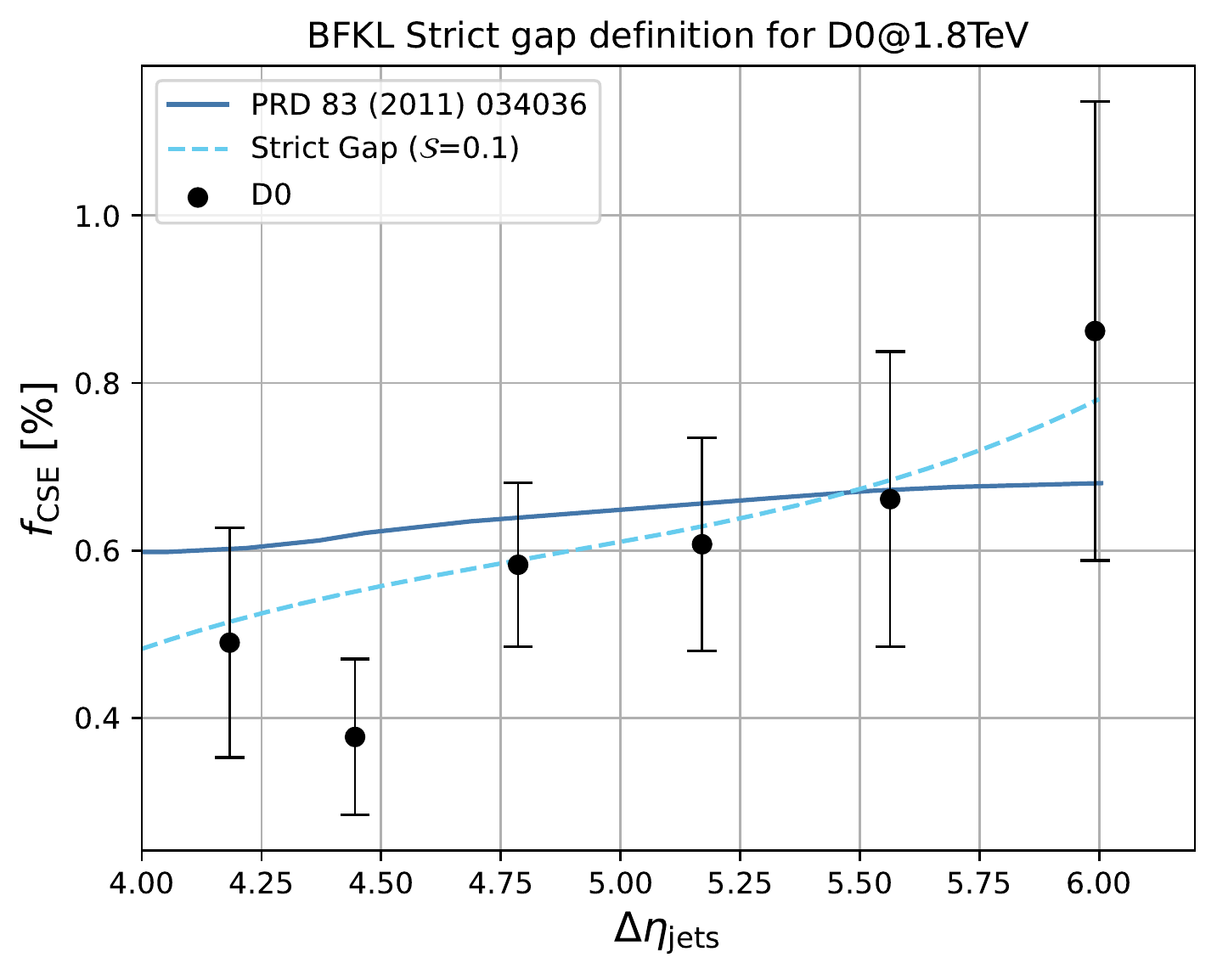}
\caption{The CSE event fraction $\fcse$ for D0 conditions \cite{Abbott:1998jb} in the range $\pTj\in[15;25]$ GeV as a function of $\Deta$ with a rapidity-ordered parton shower, no multi-parton interactions, a strict gap definition and a scale factor ${\cal S}$ of 0.1. The full blue line represents the RMK results from Ref.~\cite{Kepka:2010hu}.}
\label{fig:05}
\end{figure}

Next, we validated the $f_{\rm CSE}$ results in Fig.\ 2 of the RMK analysis \cite{Kepka:2010hu} for D0 conditions \cite{Abbott:1998jb}. Note, however, that we follow RMK here in employing a strict gap definition, i.e.\ no particles in the gap, while D0 imposed a calorimeter cut on their transverse momentum of 200 MeV and allowed for up to one particle in the gap. Since RMK employed the HERWIG6 \cite{Corcella:2000bw} parton shower, which is ordered in rapidity and is not interleaved with multi-parton interactions, we also used rapidity ordering and no MPI in PYTHIA8 (\texttt{SpaceShower:rapidityOrder = on} and \texttt{PartonLevel:MPI = off}). Both choices, but in particular the first one, reduce the emission of particles into the gap, as we observed directly in the multiplicity distribution of charged tracks in the gap, and they therefore increase the CSE event fraction. Without these choices, a significantly larger scale factor than the value fitted by RMK of $S=0.1$ would be required to describe the normalization of the D0 measurement. With these settings, we obtain a qualitatively good agreement with the RMK analysis and the D0 data, in particular in Fig.~\ref{fig:05} for the CSE event fraction $\fcse$ in the range $\pTj\in[15;25]$ GeV as a function of $\Deta$. Small differences in shape can most likely be attributed to those in the parton showers of HERWIG6 and PYTHIA8.

\subsection{Soft underlying event tunes and choice of parton densities}
\label{sec:2.4}

PYTHIA8 allows the simulation of the so-called soft underlying event activity, which consists of the beam-beam remnants and the particles that arise from MPI. These simulations are still based on pQCD, but they are quite sensitive to the rapidity ordering of the initial-state (spacelike) parton shower and the parameters used in the MPI. In particular, they introduce an impact-parameter dependence of the collision to account for the quark and gluon distributions in the proton in space. For a given hard parton–parton scattering, there is a probability of having additional scatterings given by the parton densities and LO $2\to 2$ QCD matrix elements. These squared matrix elements are divergent for $p_T \to 0 $ owing to their $1/p^4_T$ dependence. In the context of MPI, they are regulated by replacing $1/p^4_T \to 1/(p^2_T+p^2_{T,0})^2$. The quantity $p_{T,0}$ that regularizes the matrix element is directly related to the finite size of the proton and depends on the collision energy. It can be tuned to electron-positron, electron-hadron, and proton-(anti)proton data. Note that PYTHIA8 also allows to simulate single-diffractive (SD) dissociation, double-diffractive (DD) dissociation, central-diffractive (CD), and non-diffractive (ND) processes, which contribute to the inelastic cross section in hadron-hadron collisions. In SD, CD, and DD events, one or both of the beam particles are excited into color singlet states, which then decay. The SD and DD processes correspond to color singlet exchanges between the beam hadrons, while CD corresponds to double color singlet exchange with a diffractive system produced centrally.

\begin{table}[t]
\caption{PYTHIA8 tunes by the CMS collaboration \cite{CMS:2019csb}. These tunes use the Schuler-Sj\"ostrand diffraction model~\cite{Schuler:1993wr} and also include the simulation of central-diffractive (CD) processes. The number of degrees of freedom (dof) is 63.}
\label{tab:01}
\centering
\begin{tabular}{|lccc|}
\hline
PYTHIA8 parameter                   & CP1         & CP3 & CP5 \\
\hline
PDF Set                             & NNPDF3.1 LO & NNPDF3.1 NLO &  NNPDF3.1 NNLO  \\
$\alpha_s(m_Z)$                     & 0.130       & 0.118 &  0.118 \\
\texttt{SpaceShower:rapidityOrder}   &     off & off & on \\
\texttt{[MPI]:EcmRef} [GeV]                  &   7000  & 7000 & 7000 \\
$\alpha_s^\mathrm{ISR}(m_Z)$ value/order      &   0.1365/LO   & 0.118/NLO  & 0.118/NLO   \\
$\alpha_s^\mathrm{FSR}(m_Z)$ value/order      &   0.1365/LO    & 0.118/NLO  & 0.118/NLO   \\
$\alpha_s^\mathrm{MPI}(m_Z)$ value/order       &   0.130/LO     & 0.118/NLO  & 0.118/NLO   \\
$\alpha_s^\mathrm{ME}(m_Z)$ value/order       &   0.130/LO     & 0.118/NLO  & 0.118/NLO   \\
\hline
\texttt{[MPI]:pT0Ref} [GeV] & 2.4    & 1.52   & 1.41 \\
\texttt{[MPI]:ecmPow}       & 0.15   & 0.02   & 0.03 \\
\texttt{[MPI]:coreRadius}   & 0.54   & 0.54   & 0.76 \\
\texttt{[MPI]:coreFraction} & 0.68   & 0.39   & 0.63 \\
\texttt{ColorReconnection:range}              & 2.63   & 4.73   & 5.18 \\
$\chi^2$/dof                                  & 0.89   & 0.76   & 1.04 \\
\hline
\end{tabular}
\end{table}

For our predictions, we use the most recent CMS underlying event tunes CP1, CP3 and CP5 \cite{CMS:2019csb}. The corresponding parameters are listed in Table~\ \ref{tab:01}. These tunes have been fitted by the CMS experiment simultaneously to minimum-bias and underlying event activity measurements, i.e.\ charged-particle multiplicity and transverse momentum distributions, at the Tevatron and in Run-1 and Run-2 of the LHC \cite{CMS:2019csb} with NNPDF3.1 PDFs at LO, NLO and NNLO, respectively \cite{NNPDF:2017mvq}. They were validated with LEP event-shape observables as well as SD and non-SD minimum bias and underlying event observables at the Tevatron and the LHC. The fact that consistent underlying event tunes are possible at center-of-mass energies from 1.96 TeV at the Tevatron to 13 TeV at the LHC suggests that MPI can be factorized from the inclusive dijet production processes in a similar way as the gap survival probability $\mathcal{S}_\mathrm{prob}$ mentioned above for jet-gap-jet events. Note, however, that the tunes were neither fitted to nor validated with jet-gap-jet events.

In correspondence with the respective CMS extractions, the CP1 tune and NNPDF3.1 LO PDFs are used for our BFKL and LO QCD predictions, while the CP3 tune and NNPDF3.1 NLO PDFs are used for our NLO QCD predictions. As described above, initial-state radiation (ISR) rapidity ordering had to be switched on and MPI off in the BFKL cross section to obtain consistent results with the RMK study both in shape and magnitude. The static scale factors $\mathcal{S}$ are fitted to the data with a $\chi^2$ scan based on our theoretical calculations of the $f_\text{CSE}$ fractions. The uncertainties on $\mathcal{S}$ are quoted at 97.5\% CL based on these $\chi^2$ scans. We therefore keep these choices also for our comparisons to the CDF, D0 and CMS data below. However, they clearly do not agree with the PYTHIA8 CP1 tune by the CMS collaboration. For comparison, we therefore also employ the CP5 tune obtained with ISR rapidity ordering and MPI, keeping however the NNPDF3.1 LO/NLO PDFs. We have verified that the difference in PDFs is of minor importance, i.e.\ even at the LHC for $x\geq{p_T (e^{-\eta_1}+e^{\eta_2})/\sqrt{s}}={40\ {\rm GeV} \times 0.018 / 13\ {\rm TeV}}=5\times 10^{-5}$, the difference of NNLO/NLO PDFs stays below 5\%. Furthermore, in the ratio $f_{\rm CSE}$ of triple-differential cross sections, the PDFs cancel nearly perfectly (cf. Eq.\ \eqref{jgj}), as we have also verified. The main effect of the CP5 tune is therefore the dynamical simulation of the survival probability with MPI, in contrast to the additional $\mathcal{S}$ normalization factor that needs to be applied to our CP1 results. Since the CP5 tune does not necessarily model all the rescattering effects responsible for the destruction of the central gap between the jets, we supplement the calculation with a scale factor $\mathcal{S}$ as we mentioned in the previous section which is also fitted to the data with a $\chi^2$ scan. As mentioned above, it may contain effects other than MPI.

\section{Jet-gap-jet cross section}
\label{sec:3}

We now turn to our main numerical results for Mueller--Tang jet-gap-jet events. The experimental conditions of the CDF \cite{Abe:1998ip} and D0 \cite{Abbott:1998jb} experiments at the Tevatron and of the CMS experiment \cite{Sirunyan:2017rdp,Sirunyan:2021oxl} at the LHC for different center-of-mass energies $\sqrt{s}$, jet sizes $R$, subleading jet transverse momenta $p_\mathrm{T}^\mathrm{jet2}$, jet pseudorapidities $\eta_{1,2}$ ($\eta_1\eta_2<0$), gap size~$\eta^{\max}$, calorimeter/track neutral/charged particle threshold $p_T^{\min}$ and multiplicity $n^{\max}$, and observed fraction of CSE events $f_{\rm CSE}$ are listed in Table~\ \ref{tab:02}. The jet definition has evolved from cone to anti-$k_T$ clustering algorithms \cite{Cacciari:2008gp} as well as to narrower and harder jets, and the rapidity coverage has increased from 3.5 out to 4.7. A static rapidity gap has been used in all measurements, but the veto on neutral and/or charged particles was applied at different levels (calorimeter and/or tracks), and their number was allowed to vary from one to three. In this paper, we compare our BFKL NLL+PS and pQCD NLO+PS calculations to Tevatron and LHC measurements at center-of-mass energies from 1.8 to 13~TeV both in shape and normalization, paying particular attention to the effects of parton showers, multiparton interactions and the gap definition. As mentioned above, the choice of PDFs turns out to be of minor importance in the ratio $f_{\rm CSE}$ of color-singlet events. In contrast, the total scale uncertainty is dominated by the BFKL numerator, which is of ${\cal O}(\alpha_s^4(p_T))$, while the pQCD denominator is at LO only of ${\cal O}(\alpha_s^2(p_T))$, leading to a total normalization uncertainty of $\pm30\%$ for scale variations by a factor of two around the central scale $p_T$. This is only mildly altered by the NLL and NLO QCD corrections and has also been confirmed by explicit calculation.

\begin{table}[t]
\caption{Experimental conditions and results of jet-gap-jet event measurements of the CDF \cite{Abe:1998ip} and D0 \cite{Abbott:1998jb} experiments at the Tevatron and of the CMS experiment \cite{Sirunyan:2017rdp,Sirunyan:2021oxl} at the LHC for different center-of-mass energies $\sqrt{s}$, jet sizes $R$, subleading jet transverse momenta $p_\mathrm{T}^\mathrm{jet2}$, jet pseudorapidities $\eta_{1,2}$ ($\eta_1\eta_2<0$), gap size $\eta^{\max}$, calorimeter/track neutral/charged particle threshold $p_T^{\min}$ and multiplicity $n^{\max}$, and observed fraction of CSE events $f_{\rm CSE}$. The * symbol is to stress that at 7 TeV, a maximum of $n^{\max} = 3$ were used for the $p_\mathrm{T}^\mathrm{jet2} > 100 $ GeV jets.}
\label{tab:02}
\centering
\begin{tabular}{|l|rrrrrr|}
\hline
Experiment & CDF \cite{Abe:1998ip} & D0 \cite{Abbott:1998jb} & CDF \cite{Abe:1998ip} & D0 \cite{Abbott:1998jb} & CMS \cite{Sirunyan:2017rdp} & CMS \cite{Sirunyan:2021oxl} \\
\hline
\hline 
$\sqrt{s}$ [TeV] & 0.63 & 0.63 & 1.8 & 1.8 & 7 & 13 \\
\hline
$R$ & 0.7 & 0.7 & 0.7 & 0.7 & 0.5 & 0.4\\
$p_\mathrm{T, min}^\mathrm{jet}$ [GeV]    & 8 & 12 & 20 & 15   & 40 & 40 \\
                                            & -- & -- & -- & 30   & 60 & 60 \\
                                            & -- & -- & -- & --   & 100$^*$ & 100 \\
$|\eta_{1,2}|$ & [1.8;3.5] & [1.9;4.1] & [1.8;3.5] & [1.9;4.1] & [1.5;4.7] & [1.4;4.7] \\
\hline
$|\eta^{\max}|$ & 1 & 1 & 1 & 1 & 1 & 1 \\
$p_T^{\min}$ [MeV]   & 200/300 & 200/-- & 200/300 & 200/-- & --/200 & --/200 \\
$n^{\max}$     & 2/0 & 1/-- & 2/0 & 1/-- &  --/2(3$^*$) & --/2 \\
\hline
$f_{\rm CSE}$ [\%] & $2.7\pm0.9$ & $\!\!1.85\pm0.38$ & $\!\!1.13\pm0.16$ & $\!\!0.54\pm0.17$ & $\!\!0.57\pm0.16$ & $\!\!0.64\pm0.11$\\
 & -- & -- &  -- & $0.94\pm0.13$ & $0.54\pm0.13$ & $0.77\pm0.08$\\
 & -- & -- & -- &  -- & $0.97\pm0.15$ & $0.77\pm0.11$\\
\hline
\end{tabular}
\end{table}
%

\subsection{Description of CDF data at 1.8 TeV}

\begin{figure}
\centering
\includegraphics[width=0.49\linewidth]{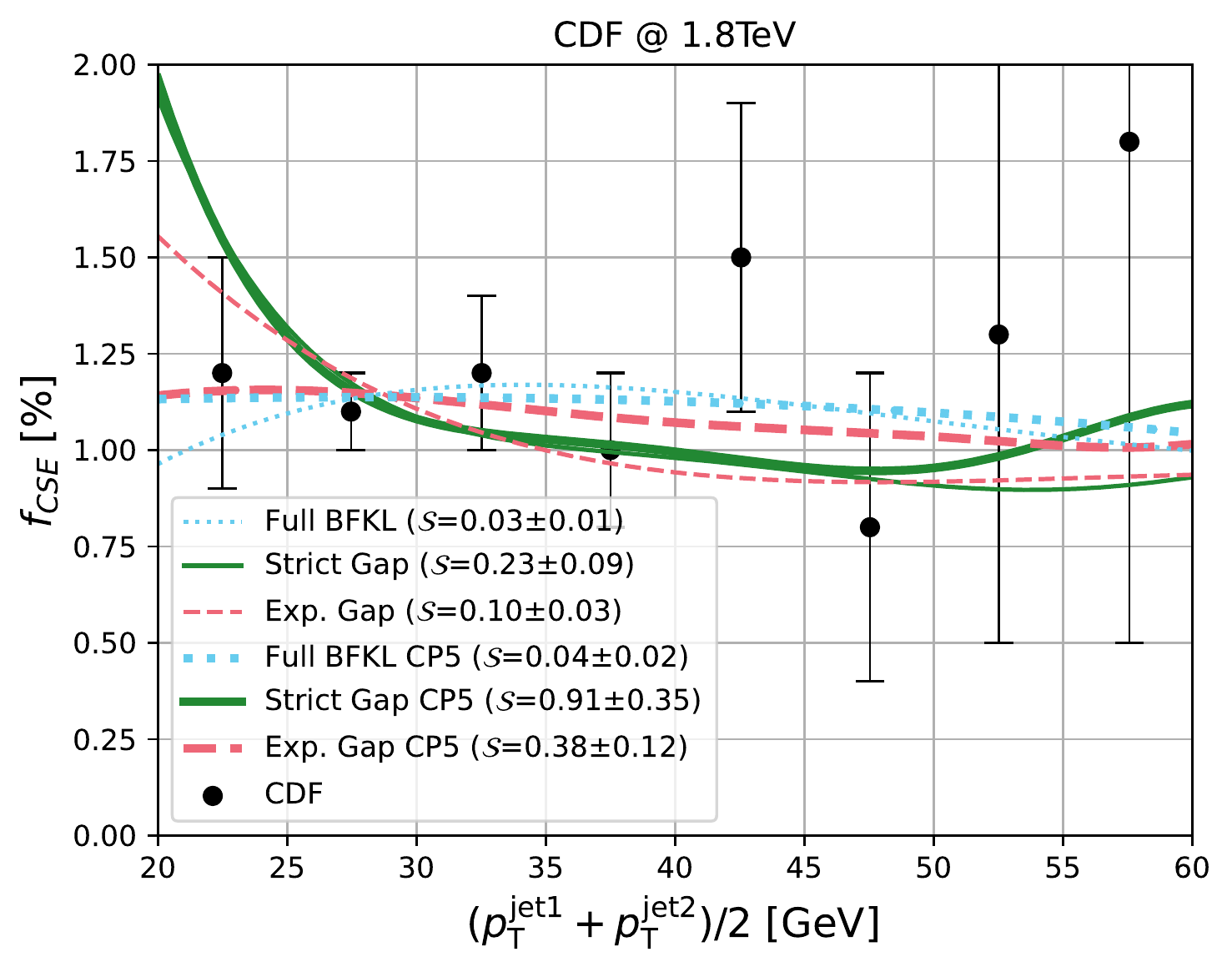}
\includegraphics[width=0.49\linewidth]{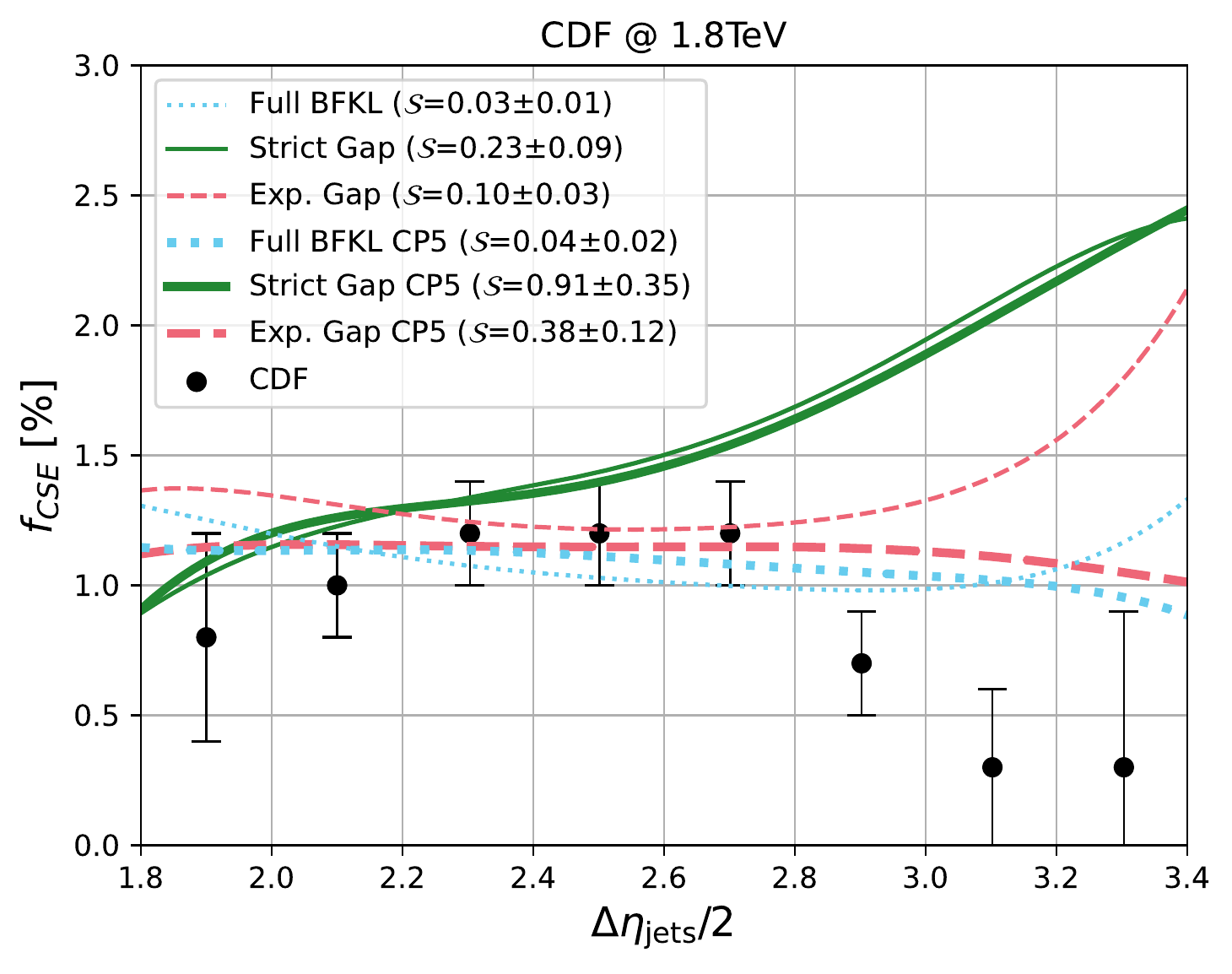}
\caption{Predictions for $\fcse$ as a function of the average jet $p_T$ (left) and $\Deta/2$ (right) compared to the CDF measurement at 1.8 TeV~\cite{Abe:1998ip} for three different gap definitions (different line shapes), using the modified tune CP1 (with rapidity ordering, without MPI, thin lines) and the tune CP5 (with rapidity ordering, with MPI, thick lines).}
\label{fig:04}
\end{figure}

We begin our comparisons in Fig.\ \ref{fig:04} with the CDF data obtained in $p\bar{p}$ collisions at a center-of-mass energy of 1.8 TeV~\cite{Abe:1998ip}. At the Tevatron, the experimental uncertainties are unfortunately of similar size as the variation in the kinematic variables average $p_T$ (left) and half pseudorapidity difference $\Delta\eta_{\rm jets}/2$ (right), which makes the identification of dynamical QCD effects difficult. This is particularly true for the CDF data.

In our theoretical predictions, we distinguish between three gap definitions:
\begin{itemize}
    \item the full BFKL prediction (light blue), where no veto on particles in the rapidity gap is applied,
    \item the strict gap definition (green), where no particles (charged or neutral) are allowed in the gap,
    \item the experimental gap definition (red), which represents an intermediate case following the experimental definition from different measurements.
    \end{itemize}
While the shapes of the theoretical $p_T$ distributions vary little with the gap definition, the scale factors obtained with the CMS tune CP1, assumed to be independent of $p_T$ and fitted to the $p_T$ spectrum, do and reflect the hierarchy of gap definitions fairly well. When all BFKL events are allowed, $S=0.03\pm0.01$; when no particles in the gap are allowed, $S=0.23\pm0.09$; and for the experimental gap definition, $S=0.10\pm0.03$. The dynamical effect of radiation into the gap is more visible in the rapidity spectrum (right), where a strict veto on particles in the gap reproduces the increase at low $\Delta\eta_{\rm jets}$, but not the fall-off at larger values, which is not reproduced by any gap definition. However, this fall-off is not very significant.

The second dynamical effect of particular interest to us is the difference of CMS tunes CP1 with no MPI (thin lines) and CP5 with MPI (thick lines). Again, the precision of the CDF data does not allow for any firm conclusions from the shapes of the distributions. However, the scale factor $S=0.38\pm0.12$ for the experimental gap definition after a dynamical modeling of the rescattering with CP5 increases significantly. While it is not consistent with one, where an additional static survival probability would no longer be required,  the statistical uncertainties from the fit are quite large. The consistency of the scale factor with unity is in fact much better for a strict gap definition ($S=0.91\pm0.35$). This might be explained by the fact that the CDF experiment has (apparently successfully) attempted to subtract the non-diffractive background to the CSE fraction.

\subsection{Description of D0 data at 1.8 TeV}

\begin{figure}
\centering
\includegraphics[width=0.49\linewidth]{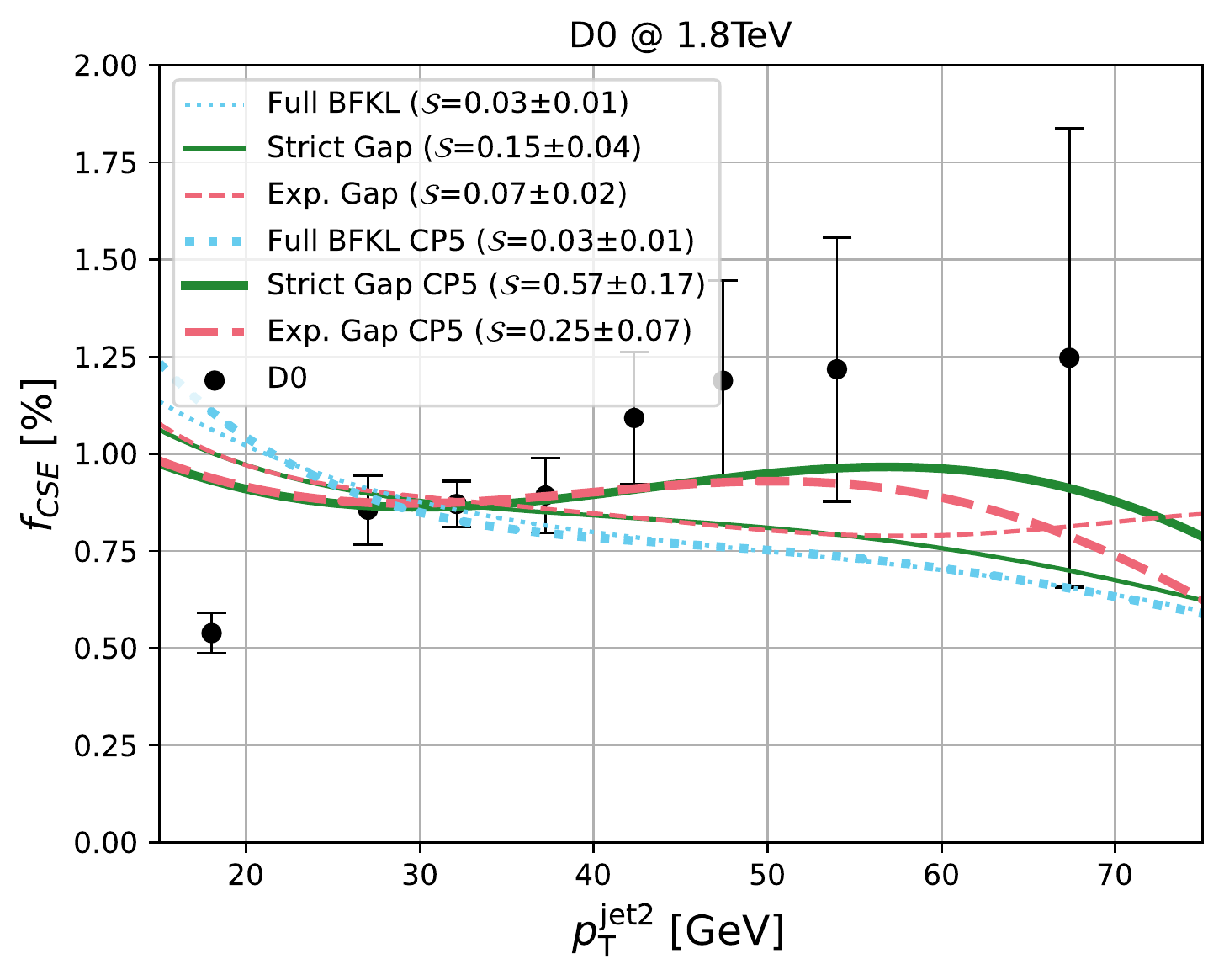}\\
\includegraphics[width=0.49\linewidth]{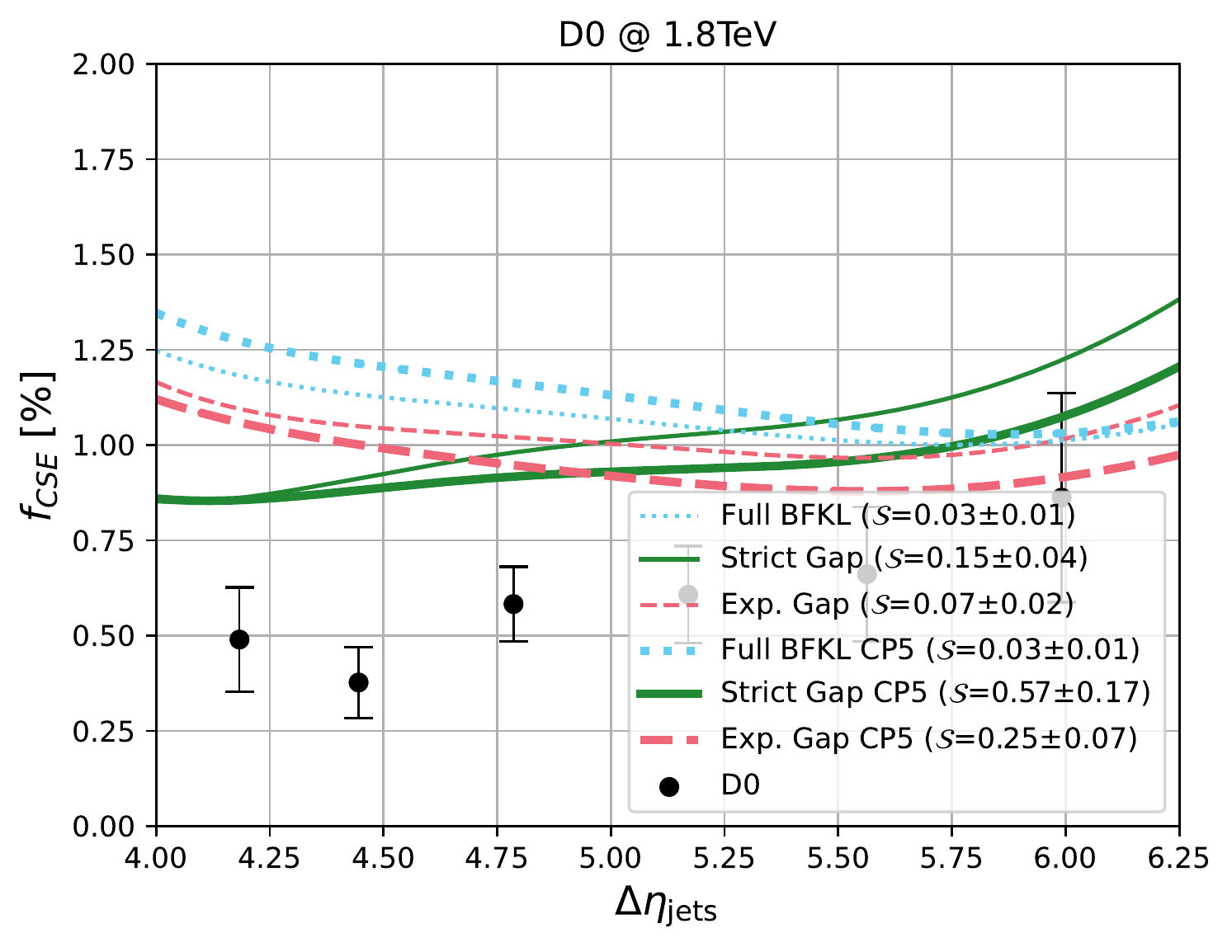}
\includegraphics[width=0.49\linewidth]{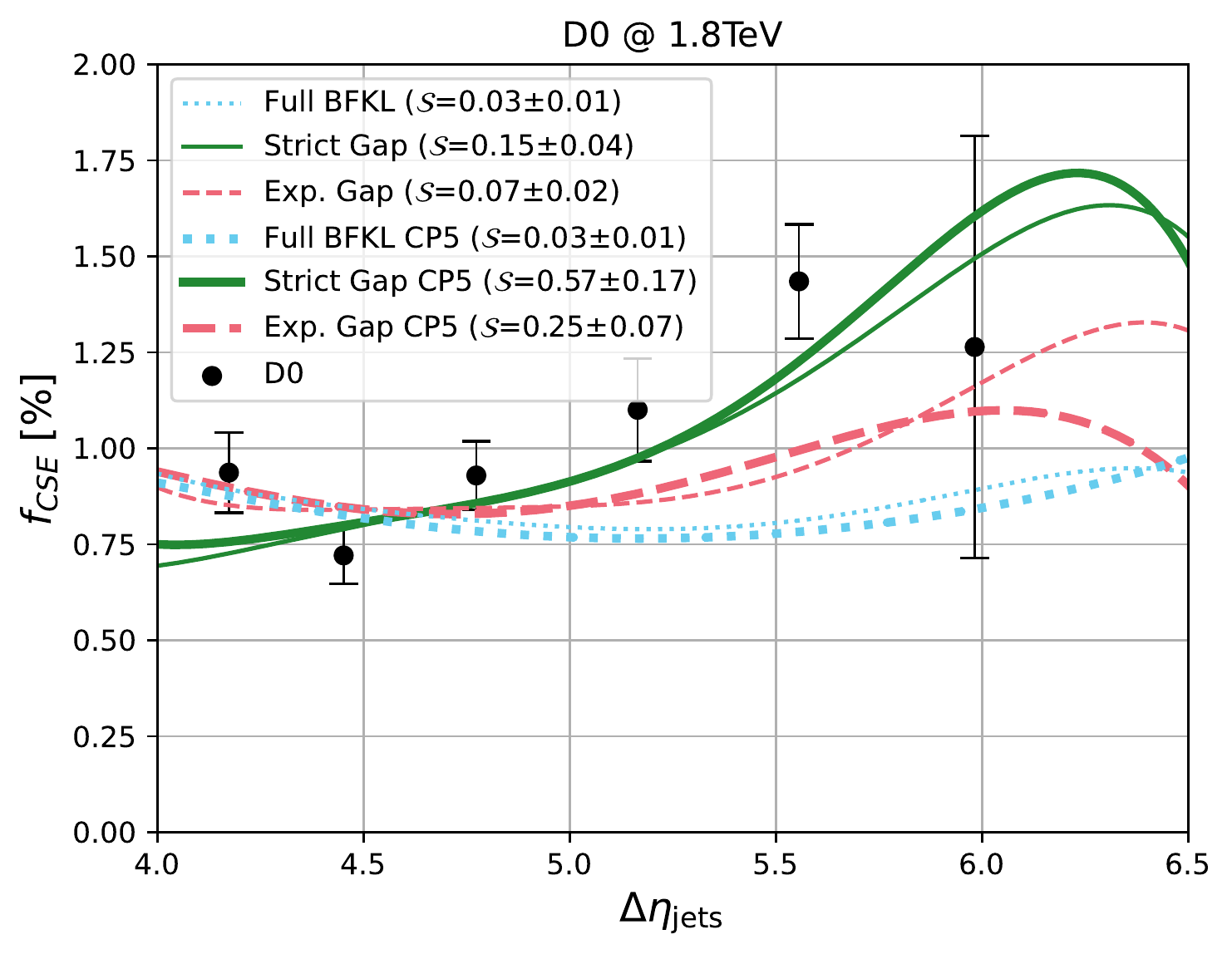}
\caption{Predictions for $\fcse$ as a function of $\pTj$ (top) and $\Deta$ with $\pTj\in[15;25]$ GeV (bottom left) and $>30$ GeV (bottom right) compared to the D0 measurement at 1.8 TeV~\cite{Abbott:1998jb} for three different gap definitions (different line shapes), using the modified tune CP1 (with rapidity ordering, without MPI, thin lines) and the tune CP5 (with rapidity ordering, with MPI, thick lines).}
\label{fig:06}
\end{figure}

Next, we compare our predictions in Fig.\ \ref{fig:06} to the D0 data, also obtained in $p\bar{p}$ collisions at 1.8 TeV center-of-mass energy \cite{Abbott:1998jb}. While the CDF data contained only 10200 events with two jets on opposite sides (OS, $\eta_1\eta_2<0$) in a single bin $p_T>20$ GeV, D0 collected more than four times as many (48000) in three different $p_T$-bins. The statistical errors are therefore smaller than in the CDF analysis and show a clear increase of $f_{\rm CSE}$ from low to high $p_T$ (top). This trend is not reproduced by the theoretical predictions irrespective of gap definition or multiple interactions. However, it hinges essentially on the lowest-$p_T$ bin, which is not well described after fitting the scale factor, the fit being dominated by the other $p_T$-bins. So for the D0 $p_T$-distribution, the differences in gap definition and CMS tune are essentially irrelevant, as they were for the CDF $p_T$-distribution.

The shapes of the D0 distributions in pseudorapidity difference are, however, well described by a strict gap definition (green), which in particular reproduces the more significant rise towards larger $\Delta\eta_{\rm jets}$ for both low $p_\text{T}^\text{jet2} \in[15;25]$ GeV (bottom left) and high $p_\text{T}^\text{jet2}>30$~GeV (bottom right). The results for medium $p_\text{T}^\text{jet2} \in[25;30]$ GeV, also measured by D0, are not shown. The fitted static scale factors (thin curves) are in agreement with those found for CDF within errors. Again the PYTHIA8 MPI (thick curves) cannot account for the full suppression of the CSE fraction, $S=0.25\pm0.07$ for the experimental gap definition being significantly smaller than one even within errors. With a strict gap definition, $S=0.57\pm0.17$ is much closer to a good dynamical description of the scale factor. Also the D0 collaboration has made an effort to subtract the background from color exchanges. This subtraction procedure dominates the systematic error in the data, as it does in the CDF analysis.

\subsection{Description of CMS data at 7 TeV}

\begin{figure}
\centering
\includegraphics[width=0.49\linewidth]{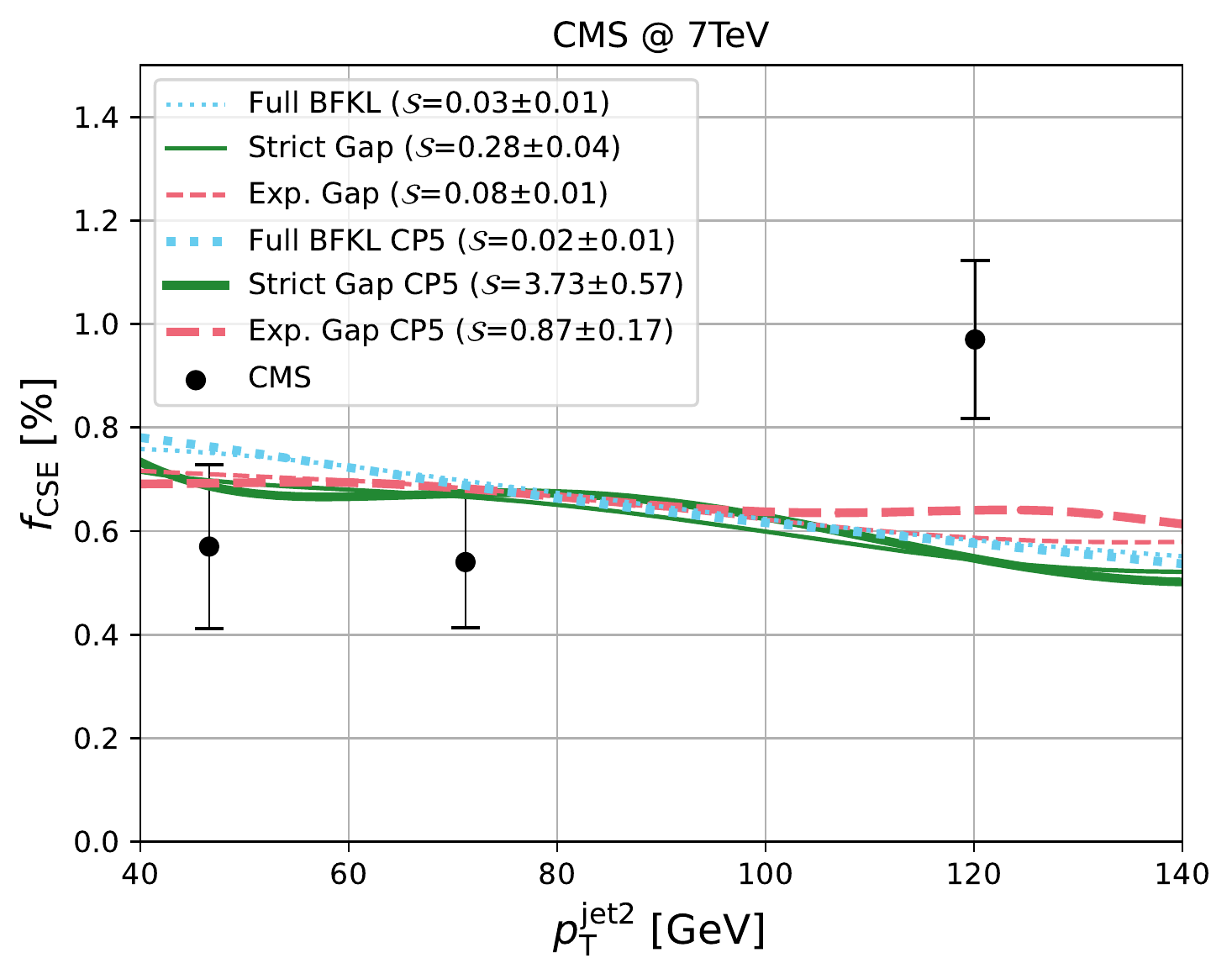}
\includegraphics[width=0.49\linewidth]{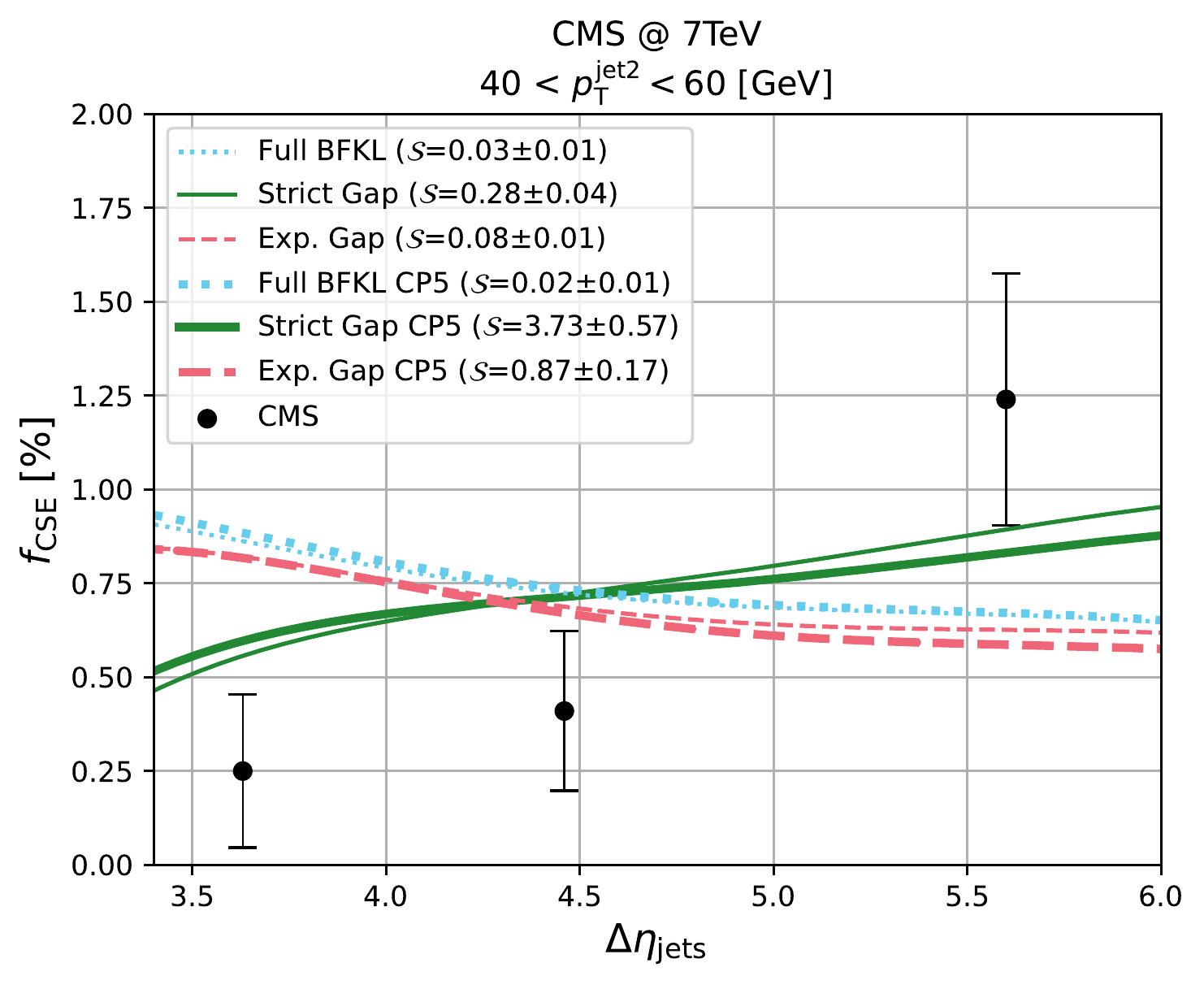}
\includegraphics[width=0.49\linewidth]{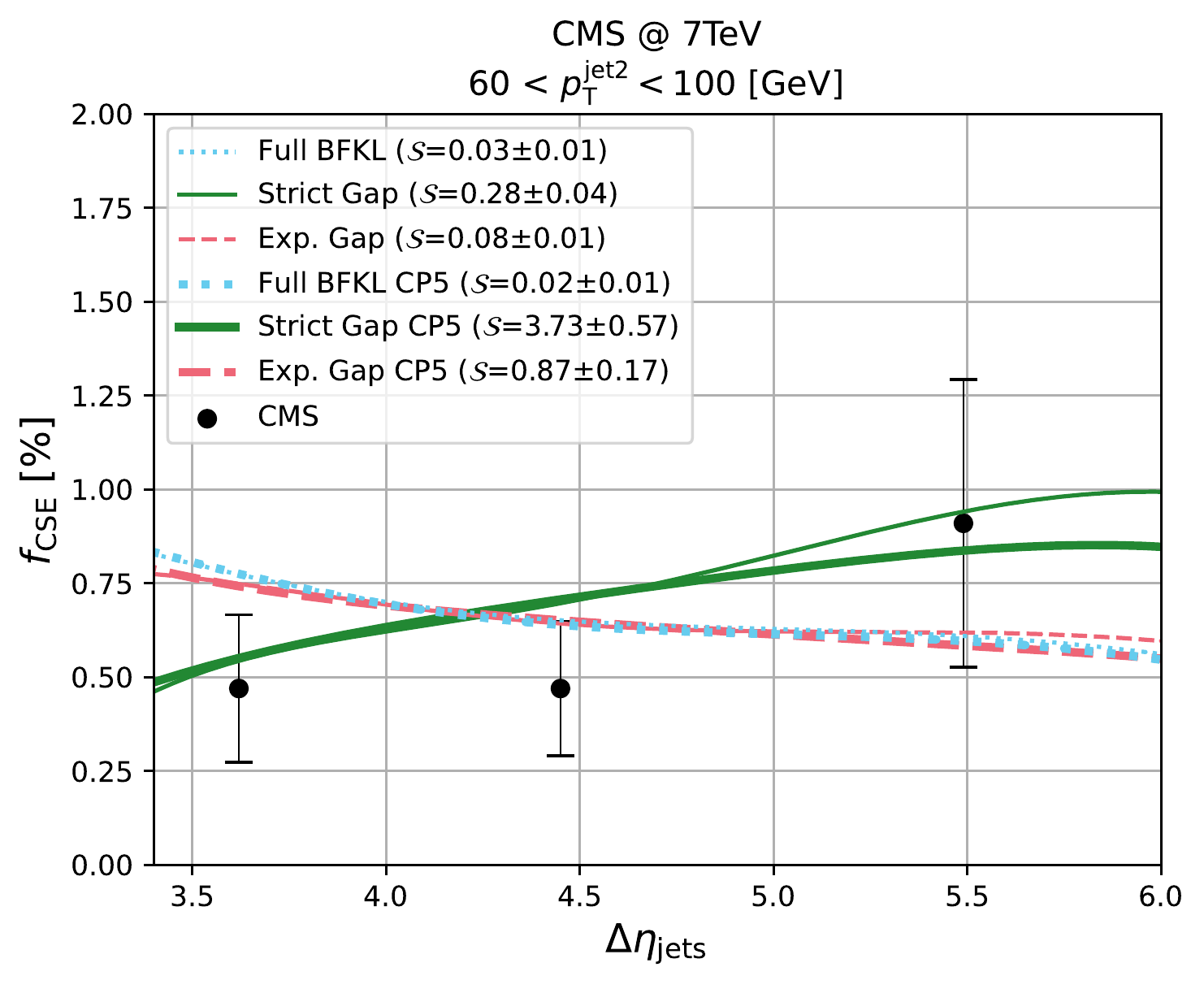}
\includegraphics[width=0.49\linewidth]{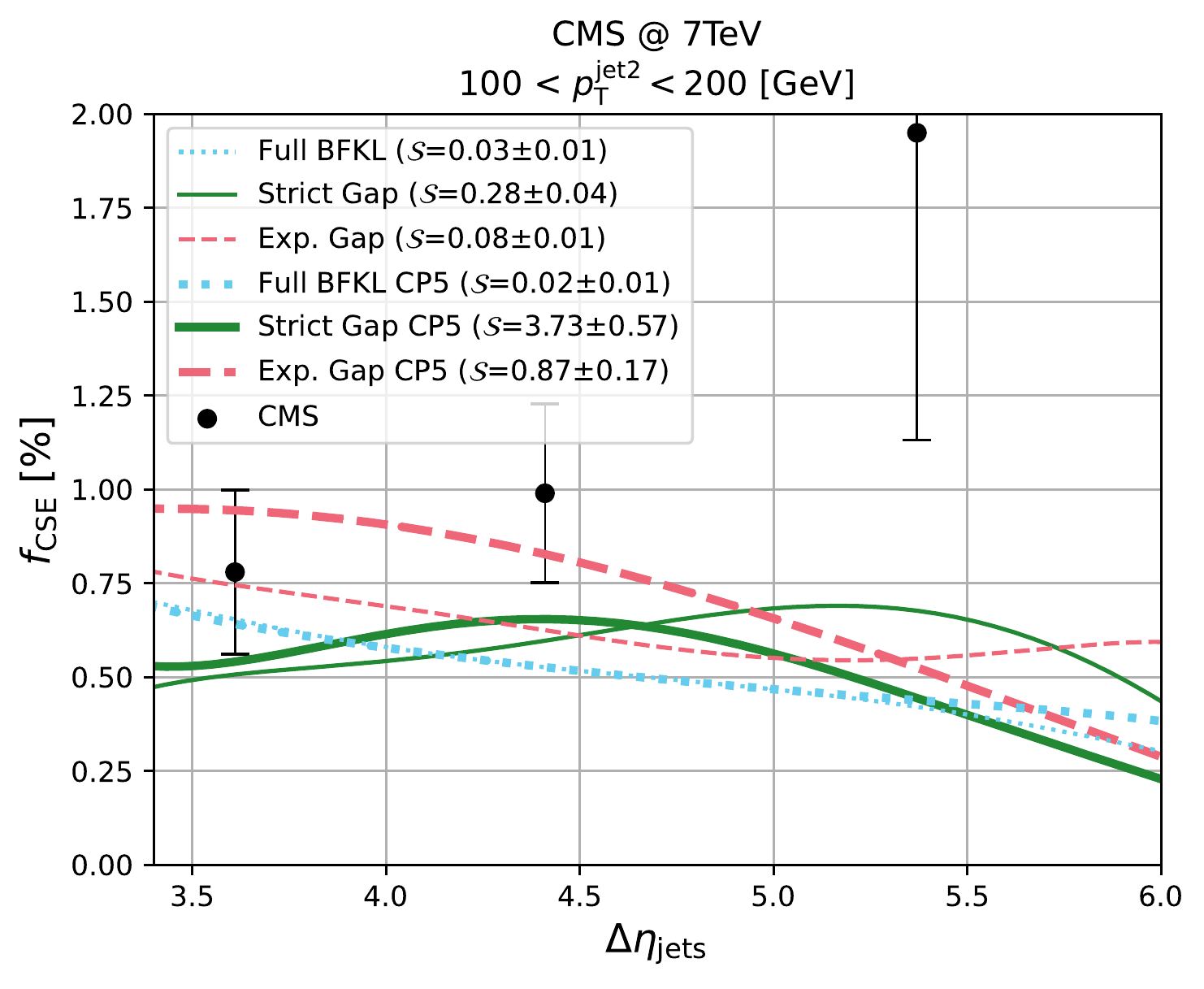}
\caption{Predictions for $\fcse$ as a function of $\pTj$ (top left) and $\Deta$ with $\pTj\in[40;60]$~GeV (top right), $\pTj\in[60;100]$ GeV (bottom left) and $\pTj\in[100;200]$ GeV (bottom right) compared to the CMS measurement at 7 TeV~\cite{Sirunyan:2017rdp} for three different gap definitions (different line shapes), using the modified tune CP1 (with rapidity ordering, without MPI, thin lines) and the tune CP5 (with rapidity ordering, with MPI, thick lines).}
\label{fig:07}
\end{figure}

We now turn in Fig.\ \ref{fig:07} to the comparisons with the CMS data in $pp$ collisions at the LHC, first at a center-of-mass energy of 7 TeV \cite{Sirunyan:2017rdp}. CMS collected 6196, 8197 and 9591 events in three different $p_T$-bins, i.e.\ for the LHC, the statistics was still relatively limited. From the second to the third-highest $p_T$-bin, the CSE fraction seems to increase slightly (top left), but the data are almost compatible with the flat theory predictions. Again, the theoretical $p_T$-distributions do not depend significantly on either the gap definition or the multiparton interactions. The shapes of the pseudorapidity distributions in the three different $p_T$-bins (top right, bottom left and bottom right) are again best described by a strict gap definition (green), the normalization showing large statistical uncertainties in the highest $p_T$-bin (bottom right) as expected from the $p_T$-distribution. For all three gap definitions, the fitted static scale factors are in agreement with those found at the Tevatron within errors~\footnote{This is only an order of magnitude since the survival probability depends on the center-of-mass energy and impact factors in the BFKL calculation also depend on center-of-mass enegies.}, i.e.\ there is no obvious center-of-mass energy dependence. The multiparton interactions in PYTHIA8 can again account very well dynamically for the suppression of CSE events with $S=0.87\pm0.17$ for the experimental gap definition.

\subsection{Description of CMS data at 13 TeV}

\begin{figure}
\centering
\includegraphics[width=0.49\linewidth]{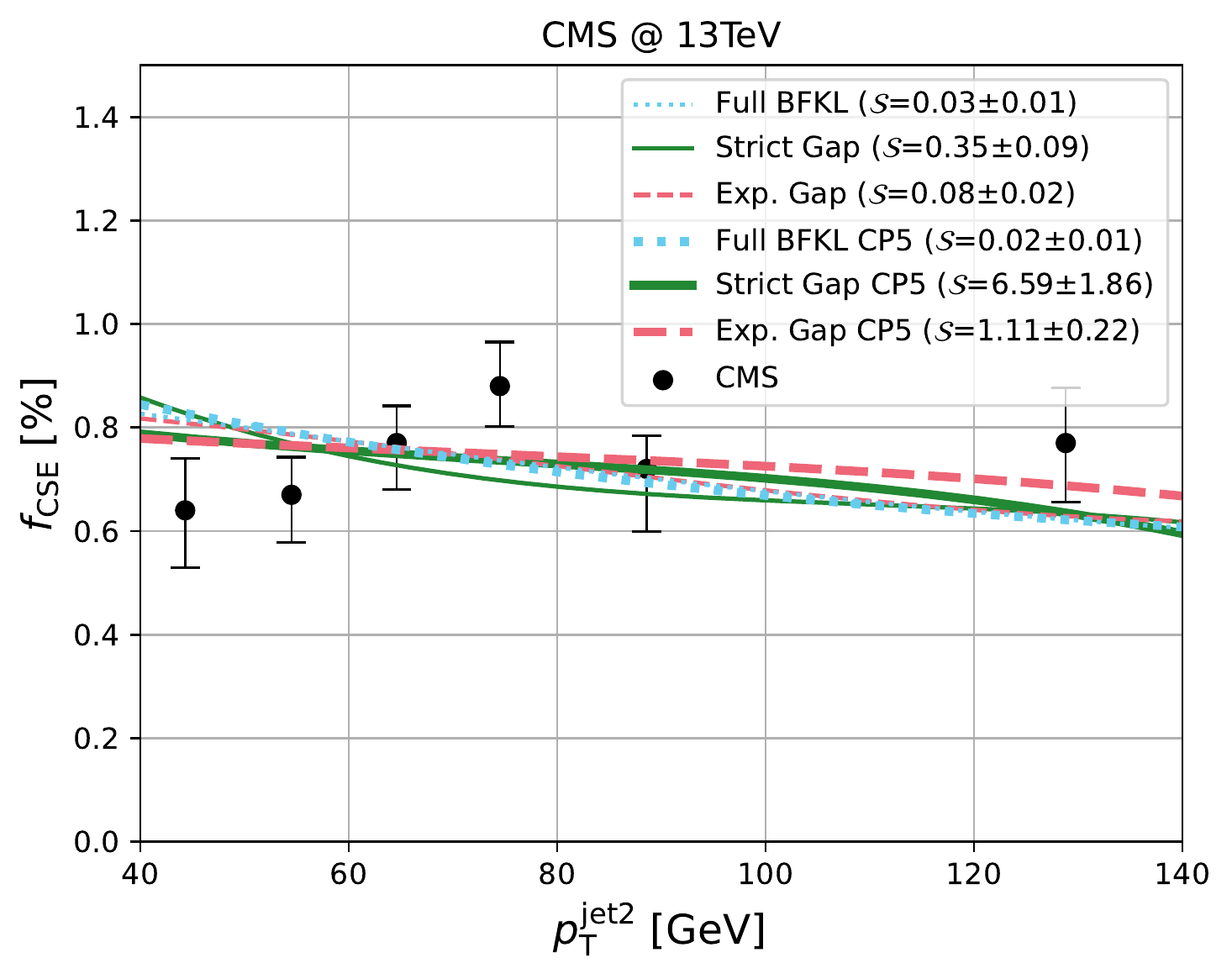}
\includegraphics[width=0.49\linewidth]{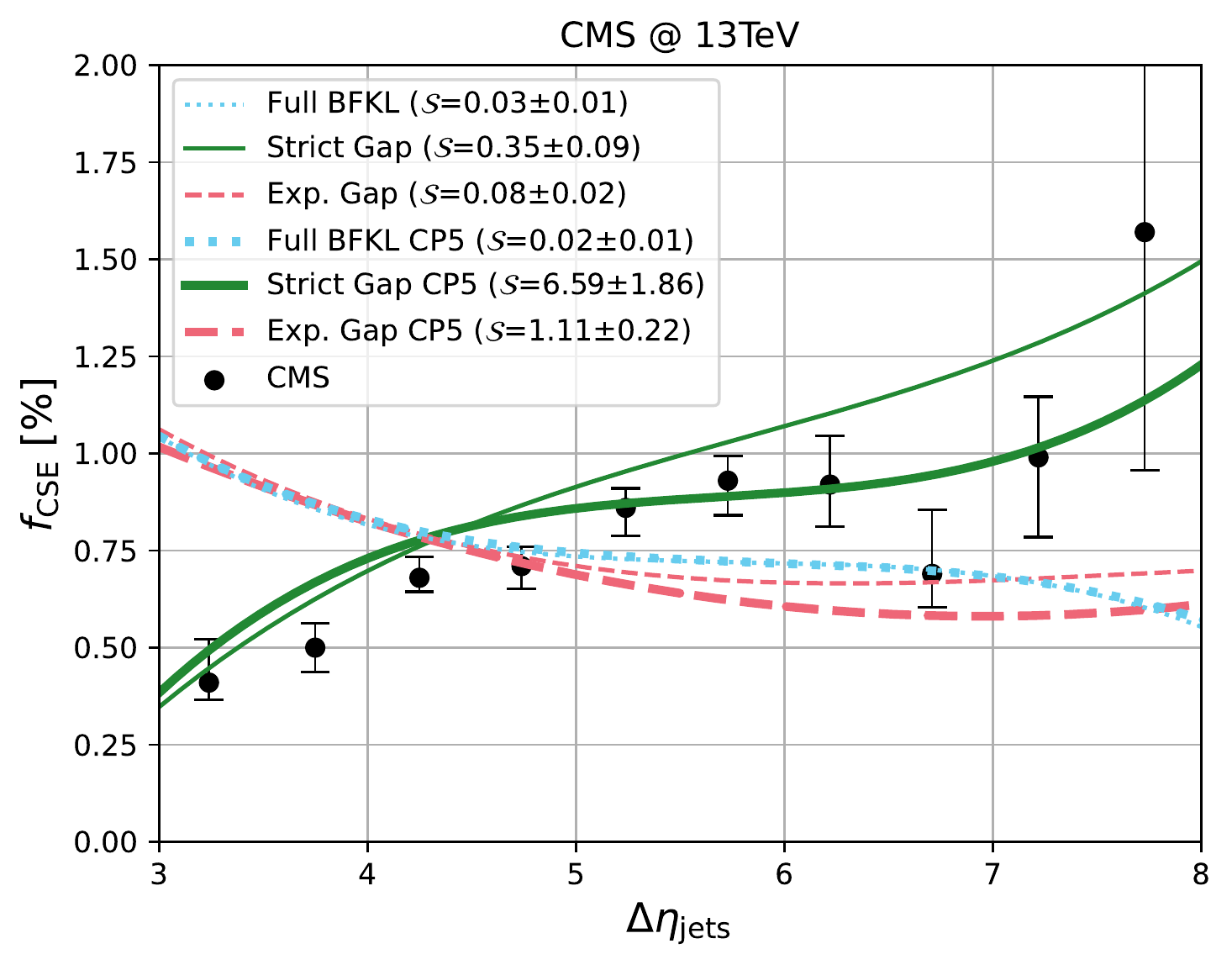}
\includegraphics[width=0.49\linewidth]{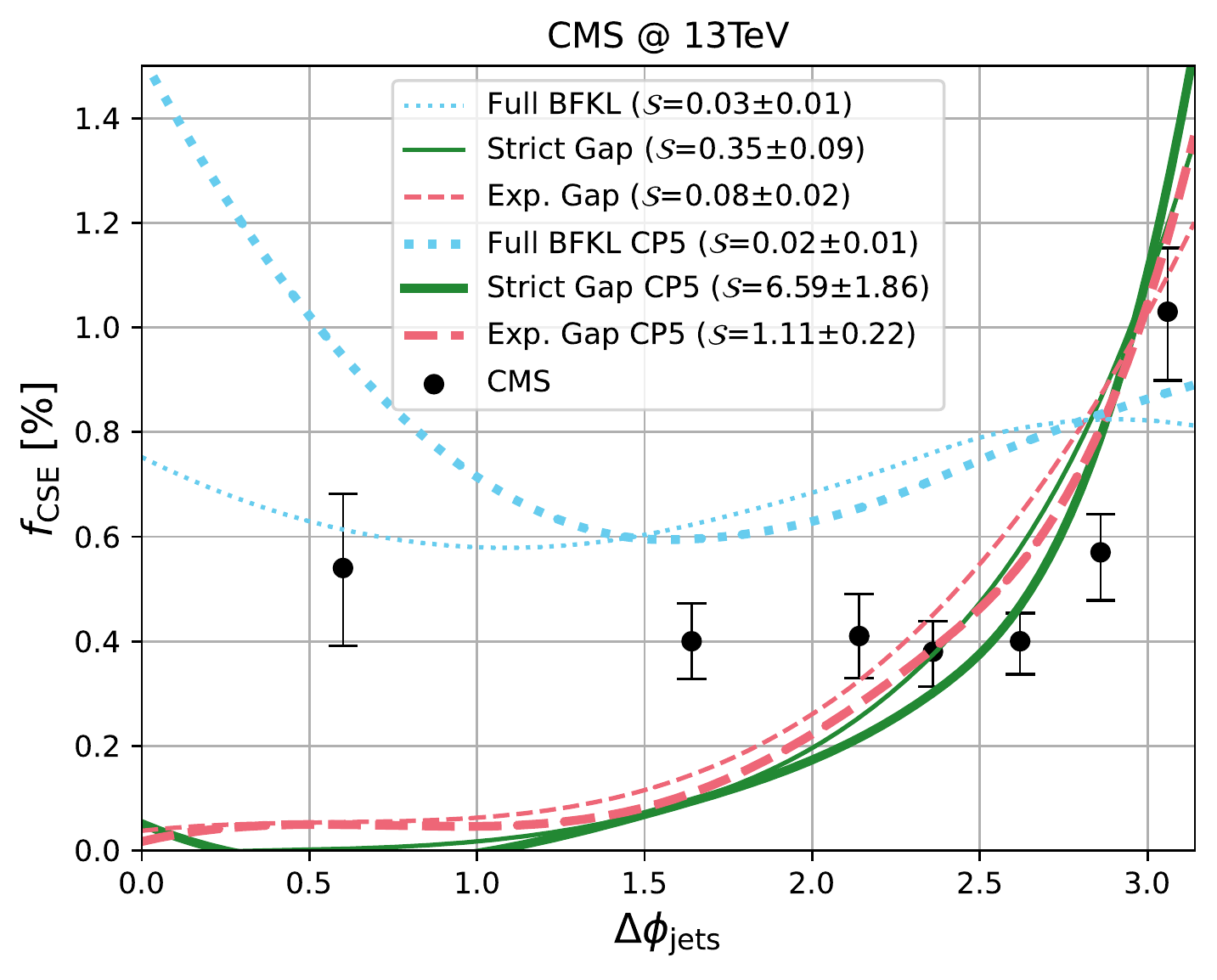}
\caption{Predictions for $\fcse$ as a function of $\pTj$ (top left), $\Deta$ (top right) and $\Dphi$ (bottom) compared to the CMS measurement at 13 TeV~\cite{Sirunyan:2021oxl} for three different gap definitions (different line shapes), using the modified tune CP1 (with rapidity ordering, without MPI, thin lines) and the tune CP5 (with rapidity ordering, with MPI, thick lines).}
\label{fig:08}
\end{figure}

Finally, we show in Fig.\ \ref{fig:08} the comparison to the CMS data at 13 TeV \cite{Sirunyan:2021oxl}. With 362915 events, this is the most precise experimental analysis so far. For the first time, not only the distribution in $p_T$ and $\Delta\eta_{\rm jets}$, but also the azimuthal angle distribution has been measured. Interestingly, the apparent rise with $p_T$ of the D0 and CMS 7 TeV data is not confirmed (top left), which is in agreement with a $p_T$-independent scale factor. Again, the theoretical $p_T$-spectra depend neither on the gap definition nor on the CMS tune. The pseudorapidity difference distribution (top right) is again best described by a strict gap definition. In contrast, only the full BFKL sample, even if probably not physical,  without any gap condition is able to describe the $\Delta\phi_{\rm jets} \equiv |\phi_\mathrm{1}-\phi_\mathrm{2}|$ distribution (bottom), i.e.\ the gap restrictions in the simulations remove too many events away from the back-to-back configuration. The theoretical normalization for both $\Delta\eta_{\rm jets}$ and $\Delta\phi_{\rm jets}$ is on the high side, as the data are dominated by the lowest $p_T$ bin, whereas the fit of the scale factor is constrained by all $p_T$-bins. The fitted scale factors are again in agreement with those obtained previously, i.e.\ even at 13 TeV there is no obvious center-of-mass energy dependence. As before, the PYTHIA8 multiparton interactions can account very well dynamically for the suppression of CSE events, $S=1.11\pm0.22$ for the experimental gap definition being consistent with one.

Let us now discuss the relevance of the gap definition that we stressed while comparing with data. In other words, we need to understand why the ``strict" gap definition leads to a better description of data, which seems to be counter-intuitive at first sight since it does not reproduce the experimental selection of events. 

\section{PYTHIA8 mechanisms for soft-particle production between jets}\label{sec:4}

\begin{figure}
    \centering
    \includegraphics[width=0.45\textwidth, page=7]{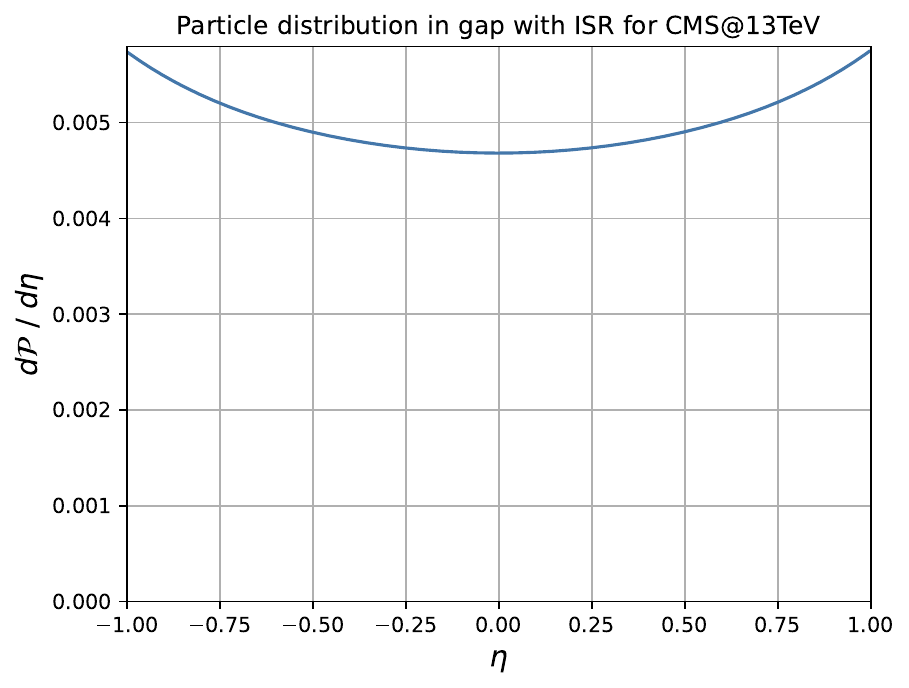}
    \includegraphics[width=0.45\textwidth, page=8]{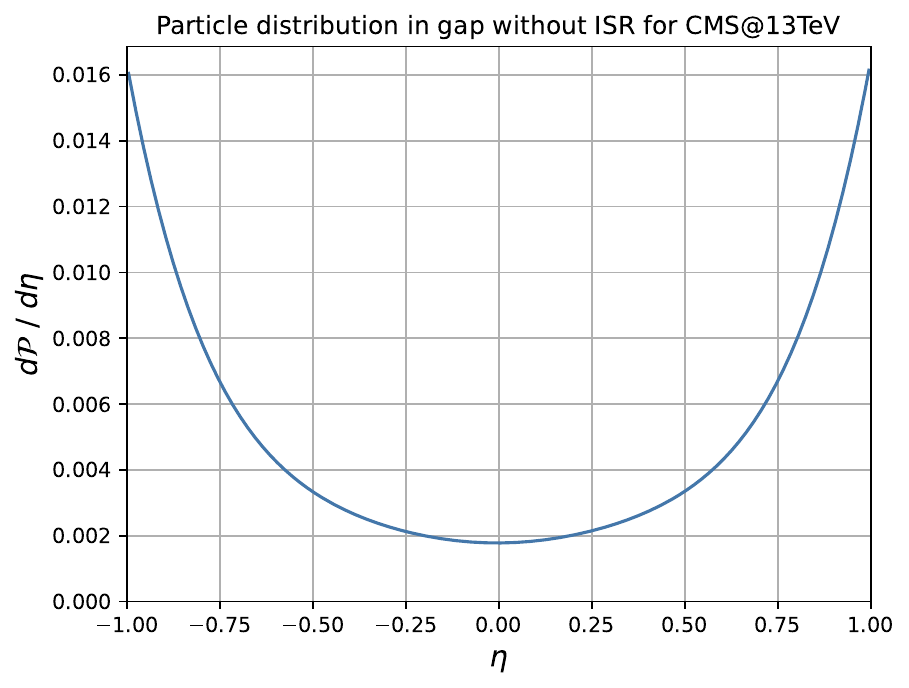}
    \caption{Distribution of charged-particles in $|\eta|<1$ between the CSE jets for \texttt{ISR = on} (left) and \texttt{ISR = off} (right). The jets satisfy the CMS selection requirements at 13 TeV. The \texttt{ISR = on} induces a flatter distribution in $\eta$. The particles that fall into the gap region for \texttt{ISR = off} are coming mostly from wide-angle unclustered hadrons from the jets.}
    \label{fig:eta_distribution_ISR}
\end{figure}

\begin{figure}
	\centering
		\begin{minipage}{0.5\textwidth}
			\includegraphics[width=1\linewidth, page=1]{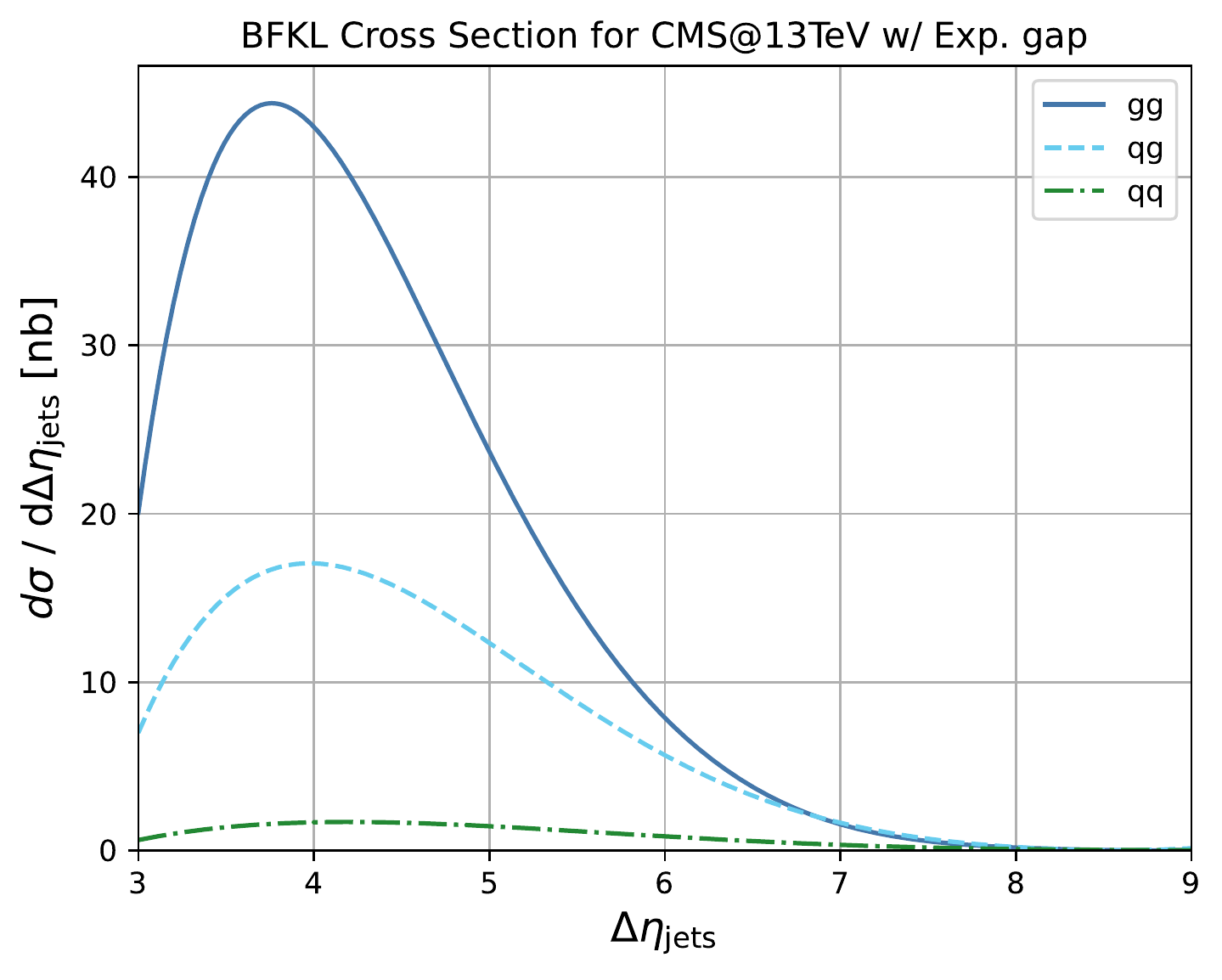}
		\end{minipage}\begin{minipage}{0.5\textwidth}
			\includegraphics[width=1\linewidth, page=1]{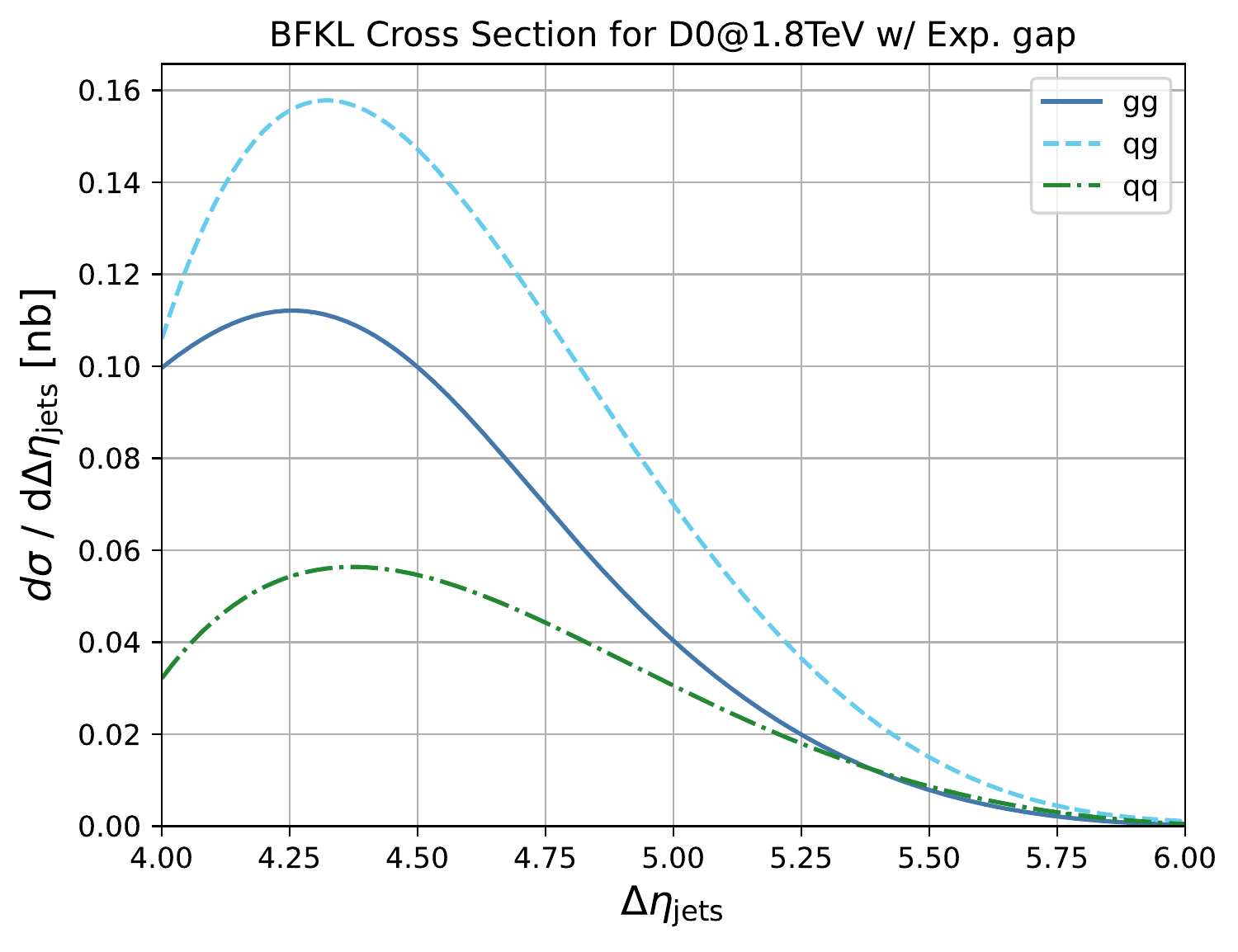}
		\end{minipage}

		\begin{minipage}{0.5\textwidth}
			\includegraphics[width=1\linewidth, page=2]{LHC_additional_plots.pdf}
		\end{minipage}\begin{minipage}{0.5\textwidth}
			\includegraphics[width=1\linewidth, page=2]{D0_additional_plots.pdf}
		\end{minipage}
    \caption{Differential cross sections as a function of $\Delta\eta_\mathrm{jets}$ for CSE dijet events. The plots on the left correspond to the CMS setup at 13 TeV, while the ones on the right correspond to the D0 setup at 1.8 TeV. The lower and upper panels correspond to the strict and experimental gaps, respectively.}
    \label{fig:dsigma_dDeltaEta}
\end{figure}

\begin{figure}
    \centering
		\begin{minipage}{0.5\textwidth}
			\includegraphics[width=1\linewidth, page=3]{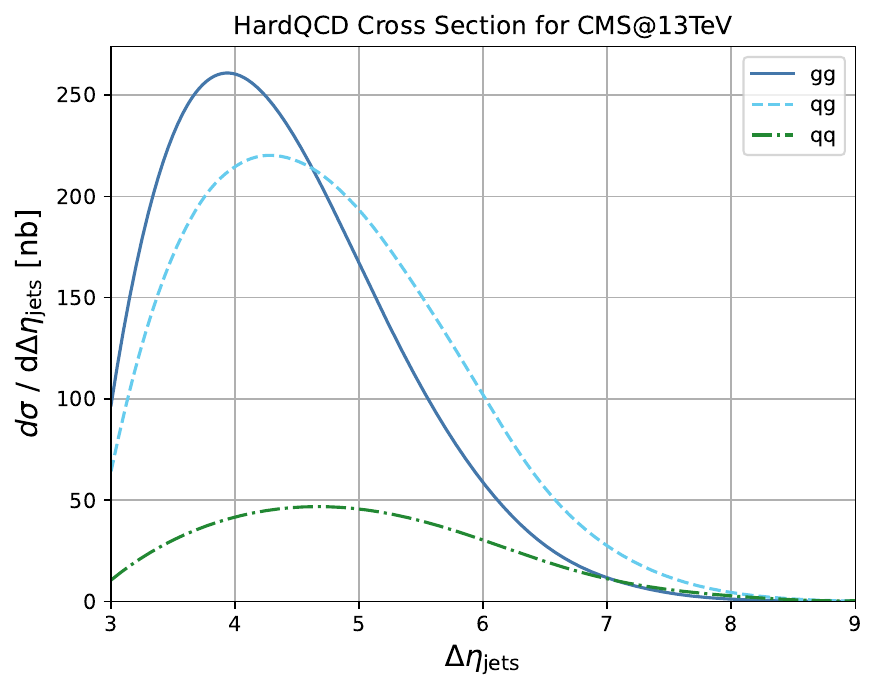}
		\end{minipage}\begin{minipage}{0.5\textwidth}
			\includegraphics[width=1\linewidth, page=3]{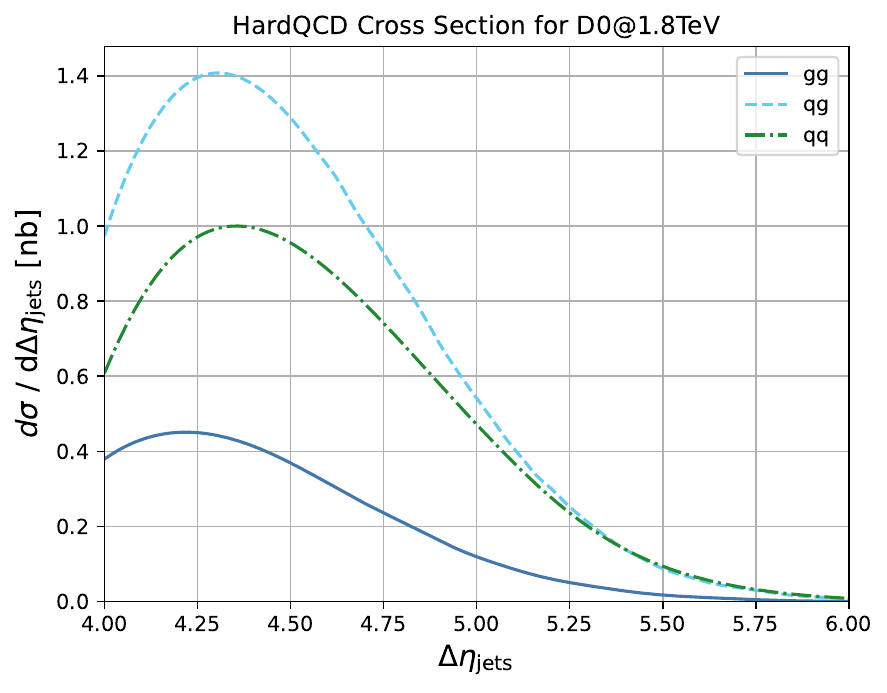}
		\end{minipage}
    \caption{QCD dijet cross section as a function of $\Delta \eta_\text{jets}$  separated in its $qq \to qq$, $gg\to gg$, and $qg\to qg$ components. The left panel is for 13 TeV collisions with the CMS selection requirements, whereas the right panel is for 1.8 TeV collisions with the D0 selection requirements.}
    \label{fig:fCSE_parton_flavour}
\end{figure}

\begin{figure}
	\centering
		\begin{minipage}{0.5\textwidth}
			\includegraphics[width=1\linewidth, page=4]{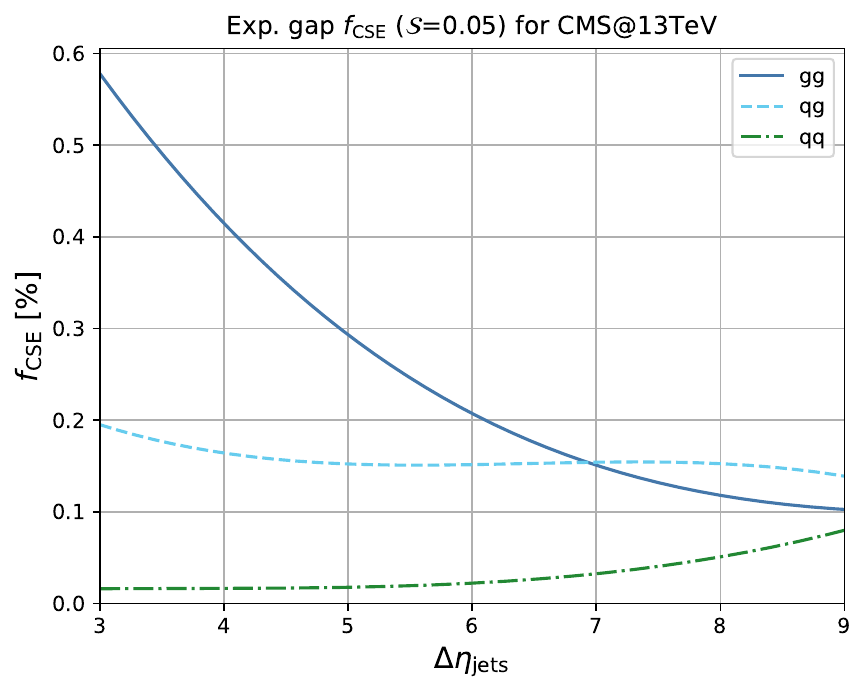}
		\end{minipage}\begin{minipage}{0.5\textwidth}
			\includegraphics[width=1\linewidth, page=4]{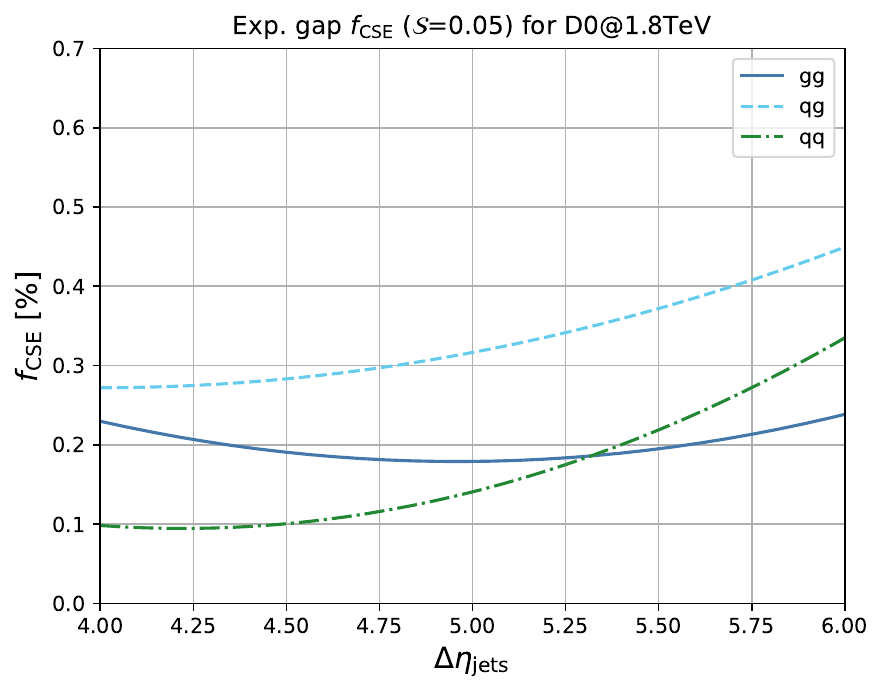}
		\end{minipage}

		\begin{minipage}{0.5\textwidth}
			\includegraphics[width=1\linewidth, page=5]{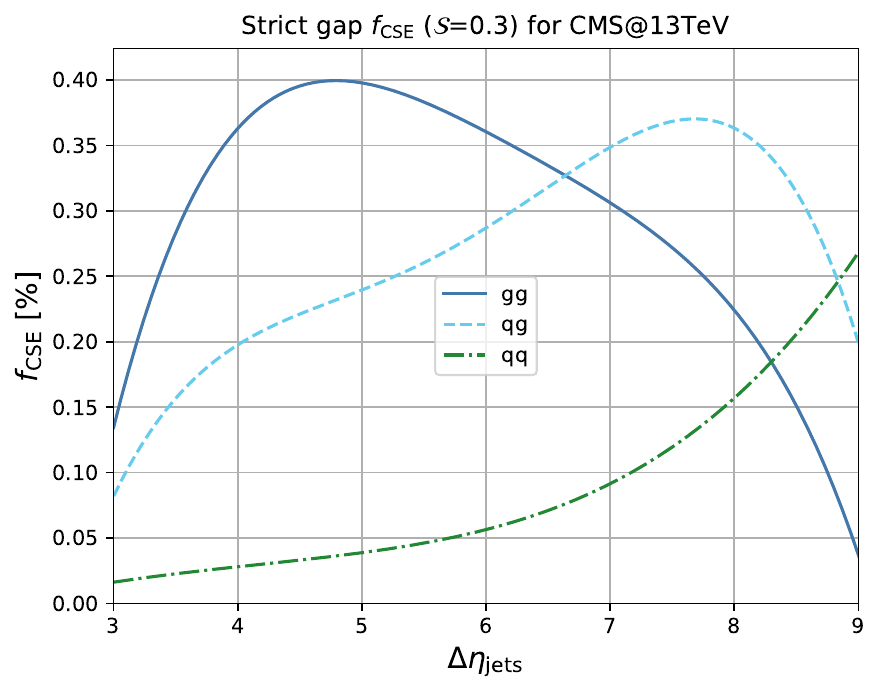}
		\end{minipage}\begin{minipage}{0.5\textwidth}
			\includegraphics[width=1\linewidth, page=5]{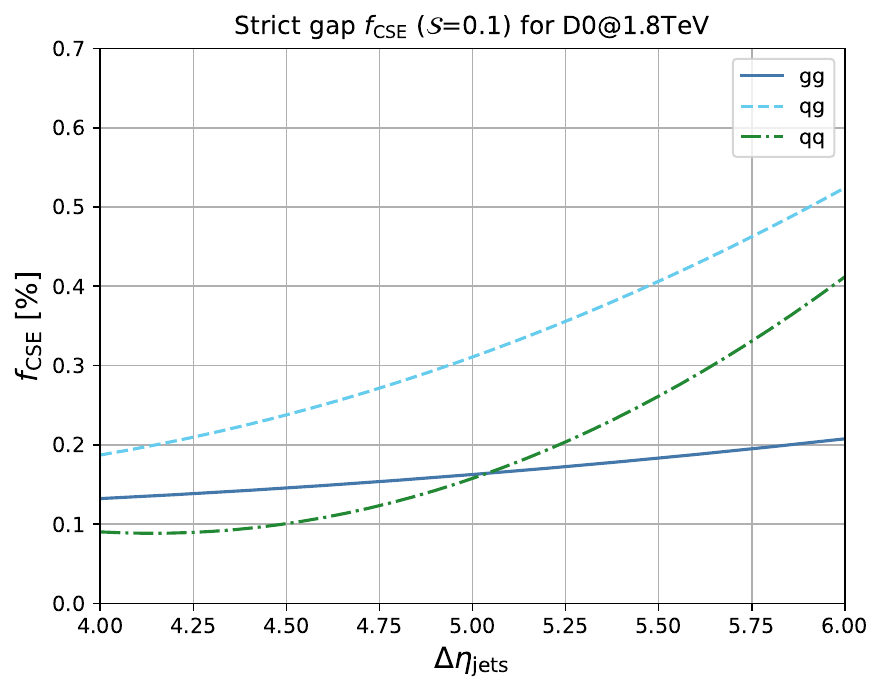}
		\end{minipage}
    \caption{Fraction $\fcse$ separated by the underlying $qq \to qq$, $gg\to gg$, $qg\to qg$ contributions to the CSE process. The two upper panels correspond to the experimental gap predictions, whereas the two lower panels correspond to the strict gap definition. The left panels correspond to the CMS 13 TeV cuts, whereas the right panels correspond to the D0 1.8 TeV cuts.}
    \label{fig:fcse_procc}
\end{figure}

\begin{figure}
	\centering
		\begin{minipage}{0.5\textwidth}
			\includegraphics[width=1\linewidth, page=9]{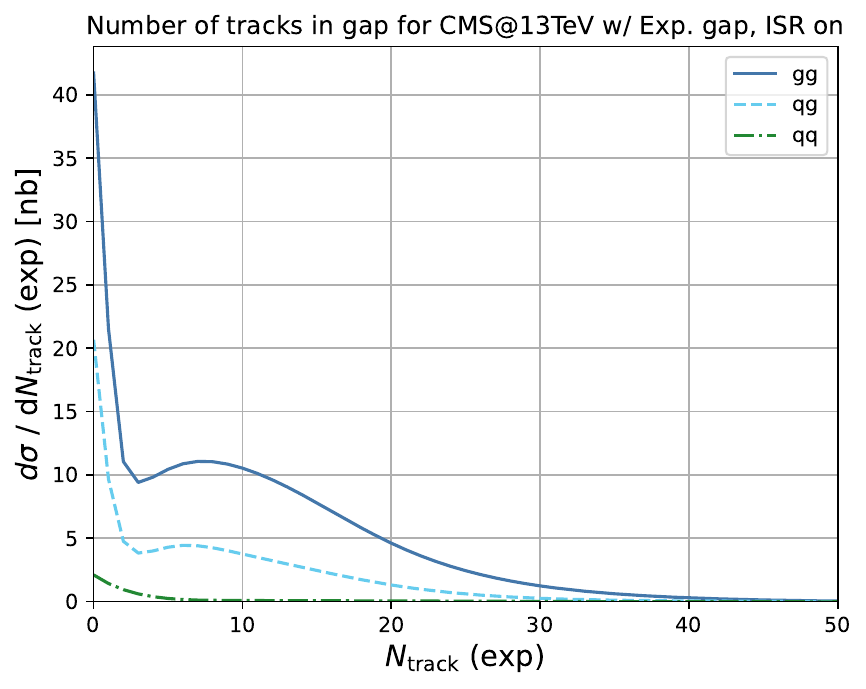}
		\end{minipage}\begin{minipage}{0.5\textwidth}
			\includegraphics[width=1\linewidth, page=10]{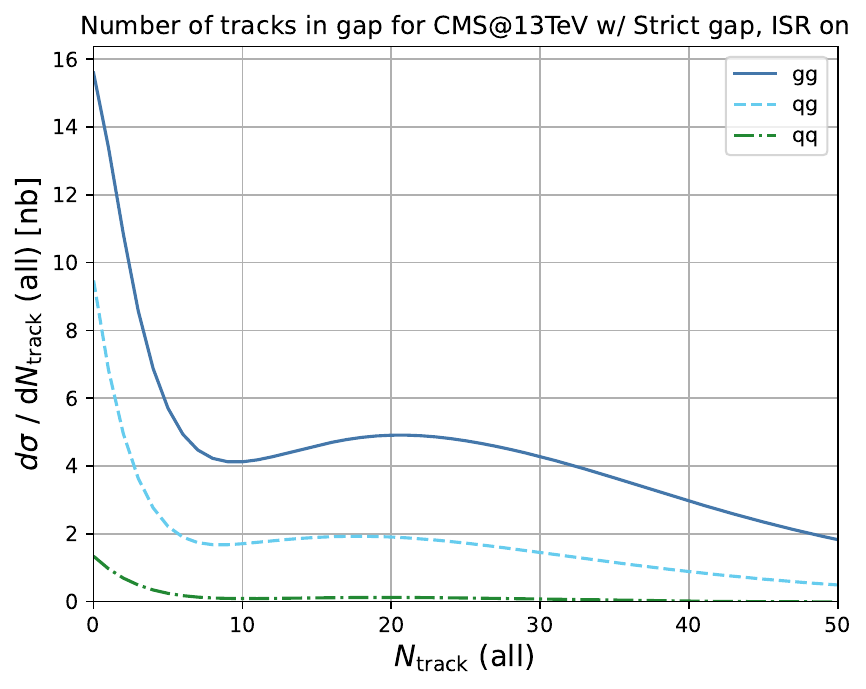}
		\end{minipage}

		\begin{minipage}{0.5\textwidth}
			\includegraphics[width=1\linewidth, page=11]{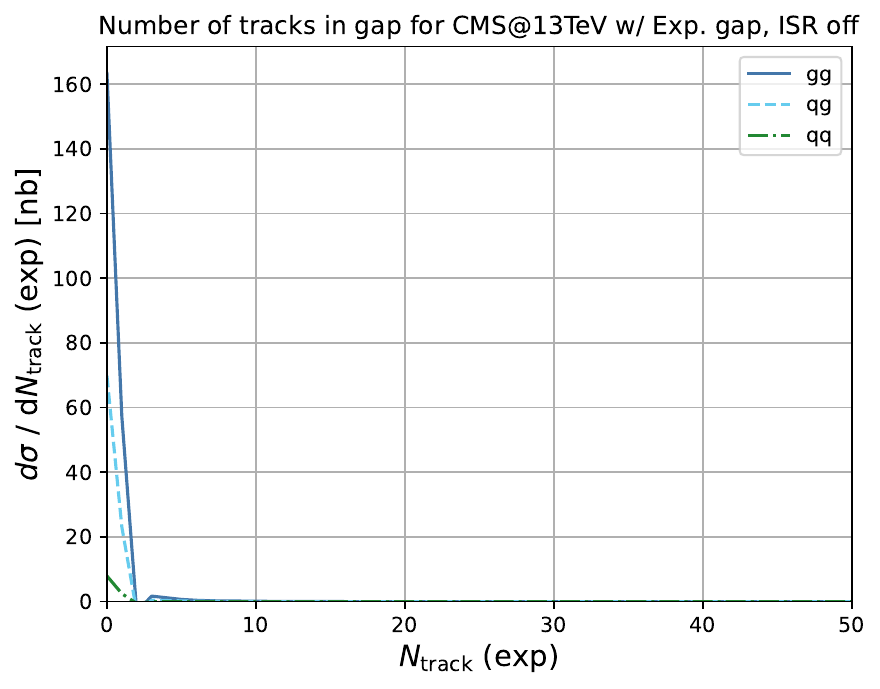}
		\end{minipage}\begin{minipage}{0.5\textwidth}
			\includegraphics[width=1\linewidth, page=12]{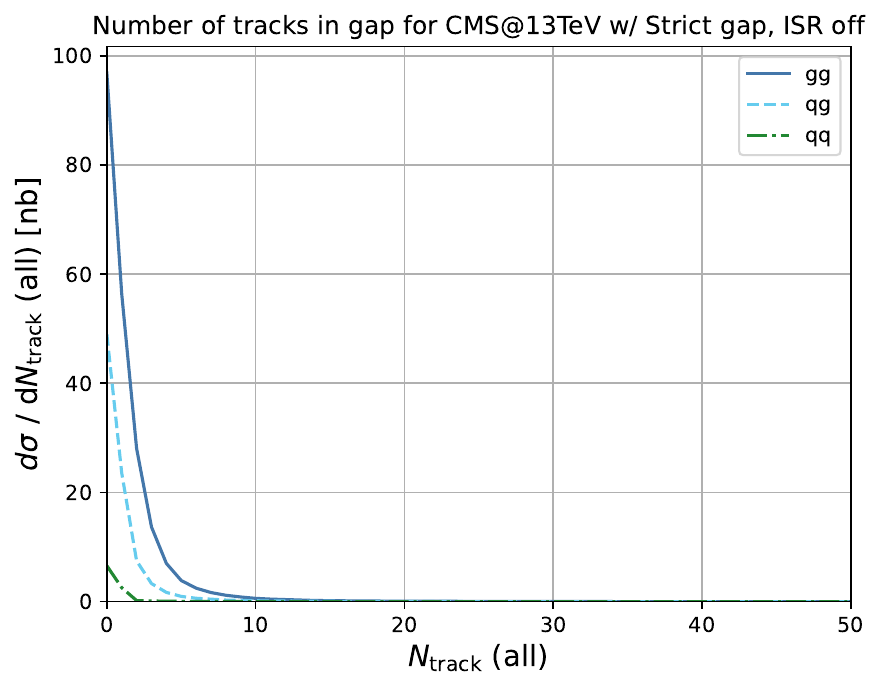}
		\end{minipage}
    \caption{$N_\text{particle}$ distributions in the $|\eta|<1$ interval between the jets. The upper (lower) left panels correspond to \texttt{ISR = on} (\texttt{ISR = off}) for an experimental gap definition, the upper (lower) right panels correspond to \texttt{ISR = on} (\texttt{ISR = off}) for the strict gap definition. The \texttt{ISR = on} induces a ``hump'' due to the breaking of additional color strings that span wide intervals in rapidity, as explained in the text. The \texttt{ISR = off} is much cleaner, since there are not as many color strings attached to the forward and backward color charges. These plots are created with the CMS @ 13 TeV analysis setup. \label{fig:Nch_ISR}}
\end{figure}

In this section, we explore the processes driving particle production between the jets in the simulation. As shown in the previous Section, the result with the strict gap yields a better description of $f_\text{CSE}$ as a function of $\Delta\eta_\text{jets}$, $p_\text{T}^\text{jet2}$ and $\langle p_\text{T}^\text{jet1,2} \rangle$, and $\Delta\phi_\text{jets}$. The most striking difference with respect to the experiment-like gap is the $\Delta\eta_\text{jets}$ dependence. It seems that we have more sensitivity to the modeling of soft particle emissions into the gap region based on the MC generator parametrization.

\subsection{Initial-state radiation and color reconnection}

We find that ISR, together with color reconnection, is the main mechanism responsible for the production of particles between the CSE dijets in our PYTHIA8 simulations. The production of multiple color charges via initial-state parton showers enhances the probability to have color strings connecting the forward-most and backward-most particle systems, effectively establishing a net color flow between the colliding protons even if there was a hard scattering with a color-singlet exchange. When these color strings eventually break, hadrons are produced between the high-$p_\text{T}$ jets, destroying the rapidity gap signature.

To understand these effects in detail, we examine the properties of CSE dijet events with \texttt{ISR = on} and \texttt{ISR = off}. In Fig.~\ref{fig:eta_distribution_ISR}, we present the particle $\eta$ distribution in the $|\eta|<1$ interval for CSE dijet events at 13 TeV. With \texttt{ISR = off}, the particle distribution peaks at the edges of the gap, suggesting that these particles are unclustered hadrons from the jets that are allocated far away from the jet axes. In contrast, the $\eta$ distribution of particles with \texttt{ISR = on} is rather flat in central $\eta$. Increasing the minimum $\eta$ of the jets relative to the gap interval (e.g., two units of $R$ instead of one unit of $R$) does not influence much the $\eta$ distributions; they remain mostly flat at central $\eta$. Thus, this effect is likely due to color reconnection.

It is instructive to decomposed the CSE dijet process into parton flavors to further characterize these effects. As mentioned previously in Section~\ref{sec:2.1}, the parton-level cross section is flavor-blind in the NLL approximation, modulo global color factors between the $gg\to gg$, $qg\to qg$, and $qq\to qq$ scatterings. However, the convolution with the PDFs and the fact that quarks and gluons radiate differently induce non-trivial parton-flavor dependence on the $f_\text{CSE}$ observable that are worth examining.

In Fig.~\ref{fig:dsigma_dDeltaEta}, we show the CSE cross section as a function of $\Delta\eta_\text{jets}$ for the strict and experimental gaps using D0 1.8 TeV and CMS 13 TeV conditions decomposed into their $gg\to gg$, $qg\to qg$, and $qq\to qq$ contributions at the hard scattering. There is a strong hierarchy in the CSE cross sections between $gg\to gg$, $qg\to qg$, and $qq\to qq$ at 13 TeV, largely in favor of $gg\to gg$. This is because of the growth of the gluon densities at 13 TeV, together with the color factor enhancement from the CSE color structure. The hierarchy is less strong for the strict gap than the experimental gap. The distributions are slightly shifted to larger $\Delta\eta_\text{jets}$ for the strict gap, as expected (jets that are slightly farther away from the gap region yield very clean gaps). In $p\bar{p}$ collisions at 1.8 TeV, $qg\to qg$ is larger than $gg\to gg$ for both the experimental and strict gaps. Also, the $qq\to qq$ contribution is almost of similar size as the $gg\to gg$. This is related to the smaller gluon densities at 1.8 TeV. Notice that the shape of the $gg\to gg$ component changes drastically with the experimental gap in contrast to the strict gap. The description of ISR for quark-initiated and gluon-initiated processes is in principle different. If the ISR is well modelled for quark-initiated, but not for gluon-initiated processes, this could lead to a worse phenomenological description at 13 TeV (dominated by gluon-initiated processes) compared to 1.8 TeV (dominated by quark-initiated processes).

For reference, we show in Fig.~\ref{fig:fCSE_parton_flavour} the cross section for LO+PS QCD jets simulated in PYTHIA8 as a function of $\Delta\eta_\text{jets}$.  The inclusive dijets are presented at LO in pQCD in order to breakdown the cross section in $2 \to 2$ processes for a more direct comparison with the CSE plots. The $gg\to gg$ and $qg\to qg$ components contribute similarly at 13 TeV. At 1.8 TeV, the $gg\to gg$ is the smallest contribution, and the cross section is dominated by the $qg\to qg$ process. The parton flavor decomposition of inclusive dijet events is important, since the main observable in the analysis is a ratio of yields.

In Fig.~\ref{fig:fcse_procc}, we show the $f_\text{CSE}$ separated into parton flavors for the strict and experimental gap for CMS 13 TeV and D0 1.8 TeV conditions. At 13 TeV, the $gg\to gg$ contribution drives the shape of $f_\text{CSE}$ significantly at small $\Delta\eta_\text{jets}$. We see that $gg\to gg$ is responsible for the decrease of $f_\text{CSE}$ for the experimental gap predictions. The $gg\to gg$ contribution is suppressed at the largest $\Delta\eta_\text{jets}$ separations, where the observable becomes dominated by $qg\to qg$. For 1.8 TeV, the $qg\to qg$ component dominates for both the strict and the experimental gap. No suppression of $gg\to gg$ is observed at large $\Delta\eta_\text{jets}$.

In Fig.~\ref{fig:Nch_ISR}, we show the multiplicity of particles between the CSE jets for the strict and experimental gap using \texttt{ISR = on} and \texttt{ISR = off}. ISR effects induce a ``hump'' in the multiplicity distribution for $gg\to gg$ scattering for the experimental and strict gap. It results in a smearing effect for $qg\to qg$ and $qq\to qq$ for the experimental gap. This is because gluons carry two color charges, which only increases the likelihood that there are more color charges produced in ISR and FSR and  color strings connecting them, spanning wide intervals in rapidity. Notice that for \texttt{ISR = off}, the multiplicity distributions are sharply concentrated close to 0 for either the strict or experimental gap definitions.

To conclude this study, it is clear that the gap definition is very sensitive to ISR as it is for instance implemented in PYTHIA8. It seems that ISR is too large to reproduce the observation of jet-gap-jet events in data and it should be further tuned to data.

\subsection{Multiparton interactions and rapidity gap survival probability}

In this section, we study other parameters that might modify the prediction of jet-gap-jet ratios, namely MPI and their effect on the gap survival probability. 

The gap survival probability $\mathcal{S}_\mathrm{prob}$ quantifies the suppression of the CSE cross section due to the destruction of the gap between the jets. The destruction of the gap is due to the low momentum transfer interactions that occur in addition to the hard CSE scattering. The survival probability is difficult to calculate from first principles and in general is process-dependent. A component of the survival probability is due to MPI, which can be investigated in MC simulation.

\begin{figure}
    \centering

		\begin{minipage}{0.5\textwidth}
			\includegraphics[width=1\linewidth, page=6]{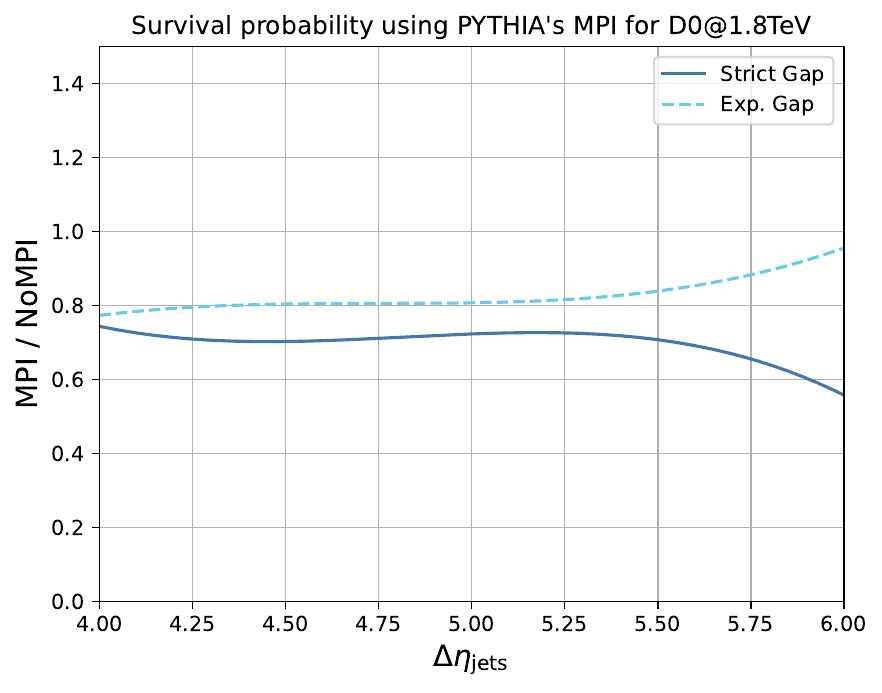}
			
		\end{minipage}\begin{minipage}{0.5\textwidth}
			\includegraphics[width=1\linewidth, page=6]{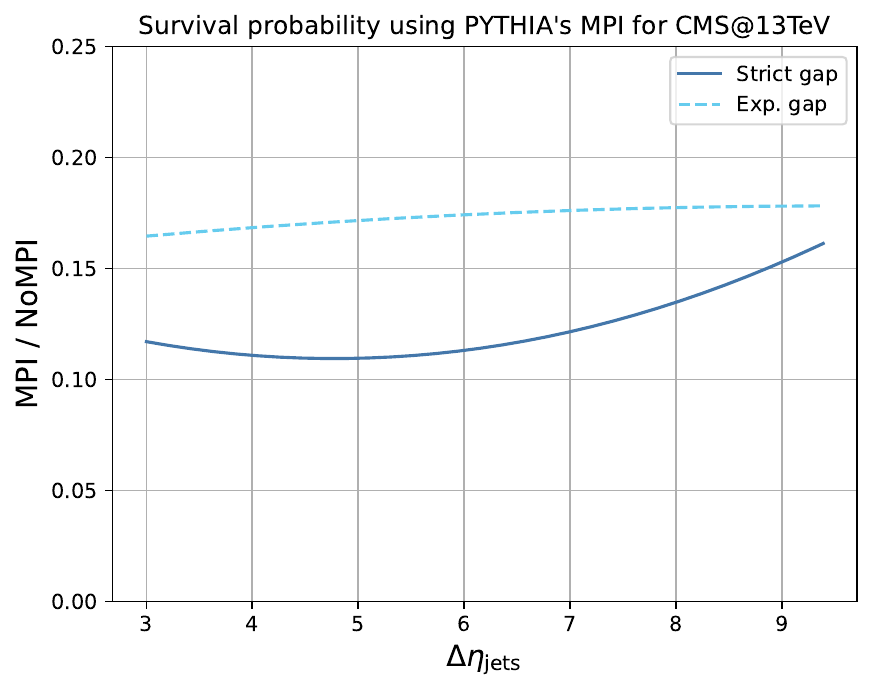}
		\end{minipage}
    \caption{Proxy for the survival probability calculated with the MPI of PYTHIA8 as a function of $\Delta\eta_\text{jets}$ between the jets. The jets on the left and right panels satisfy the D0 1.8 TeV and CMS 13 TeV dijet selection requirements, respectively. The solid blue curve shows the prediction for the strict gap, whereas the dotted-dashed cyan curve shows the prediction for the experimental gap. The tune CP1, \texttt{ISR = on} and \texttt{SpaceShower:rapidityOrder = on} settings are used for these predictions.}
    \label{fig:SP}
\end{figure}

We calculate a proxy for the survival probability assuming it originates from MPI in our PYTHIA8 simulation. The proxy for the survival probability is calculated as
\begin{equation}
    \mathcal{S}^\text{MPI}_\mathrm{prob} = \frac{f_\text{CSE}[\texttt{MPI = off}]}{f_\text{CSE}[\texttt{MPI = on}]}
\end{equation}
The resulting $\mathcal{S}^\text{MPI}_\mathrm{prob}$ as a function of $\Delta\eta_\text{jets}$ is shown in the left and right panels of Fig.~\ref{fig:SP} for the 1.8 and for 13 TeV setups, respectively. No strong dependence with $\Delta\eta_\text{jets}$ is observed, consistent with the usual assumption that $\mathcal{S}_\mathrm{prob}$ can be fitted to data as a global scale factor. The values of $\mathcal{S}^\text{MPI}_\mathrm{prob}$ are 10\% and 17\% for the strict and experimental gap definitions at 13 TeV, respectively. The reduction of $\mathcal{S}^\text{MPI}_\mathrm{prob}$ with $\sqrt{s}$ fits our qualitative expectations.

In the study in Ref.~\cite{Babiarz:2017jxc}, it was suggested that the MPI-based survival probability should increase eventually for large $\Delta \eta_\text{jets}$, since in such kinematic configuration there should not be as much energy available for MPI to take place (see Figure 5c of Ref.~\cite{Babiarz:2017jxc}, for a fixed rapidity of the dijet system $y_\text{jj} = 0$ and a fixed dijet mass $M = 200$ GeV and $p_\text{T} > 40$ GeV for the Mueller--Tang jets). We do not observe such an enhancement at large $\Delta\eta_\text{jets}$, most likely due to the difference in event selection requirements (the dijet mass and rapidity are not fixed in our case, for example). In addition to MPI, one could have contributions of soft-color interactions, where soft color-charge exchanges may rearrange the overall color structure of the collision. Also according to the calculations in Refs.~\cite{csp,cspLHC}, it was expected that the survival probability should increase with larger $\Delta\eta_\text{jets}$, but this is not observed with our calculations based on the default settings of color reconnections of PYTHIA8.

In conclusion, it seems that the MPI and survival probability effects on the shapes of the distributions are subleading with respect to the effect of ISR as mentioned in the previous section.

\subsection{Rapidity gap definition ($p_{\rm T}$ threshold, particle species, particle multiplicities)}

The electric charge of the particles used in the definition of the gap does not affect the shape $f_\text{CSE}$, as expected from the approximate isospin symmetry of the strong interactions. We also investigated whether counting exactly 0 particles, or allowing up to 1 or 2 particles in the gap affected the shape of $f_\text{CSE}$. We found that this does not have a significant impact on the shape of $f_\text{CSE}$ either, although it does on the normalization. Thus, the $p_\text{T}$ threshold is by-and-large the most important ingredient for the definition of the gap, and it seems to be directly sensitive to the fragmentation model and to the production of color charges by ISR, according to PYTHIA8 predictions.

\section{Scale factors $\mathcal{S}$}
\label{sec:5}

\begin{table}[t]
\caption{Fitted scale factors $\mathcal{S}$ and predicted survival probabilities $\mathcal{S}_\mathrm{prob}$ for different experimental conditions, center-of-mass energies $\sqrt{s}$ and gap definitions. The fit is done via a $\chi^2$ scan of the survival probabilities. The uncertainties we quote correspond to the 97.5\% confidence interval derived from the $\chi^2$ parameter scan. Assuming that the additional factors between $\mathcal{S}$ and $\mathcal{S}_\mathrm{prob}$ show a smooth dependence on kinematics and $\sqrt{s}$, we can study the relative difference between both results.}
\label{tab:03}
\centering
\begin{tabular}{|l|rrrr|}
\hline
Experiment & CDF \cite{Abe:1998ip} & D0 \cite{Abbott:1998jb} & CMS \cite{Sirunyan:2017rdp} & CMS \cite{Sirunyan:2021oxl} \\
\hline
\hline
$\sqrt{s}$ [TeV] & 1.8 & 1.8 & 7 & 13 \\
\hline
Full BFKL \hfill PYTHIA8     & 0.03$\pm0.01$ & 0.03$\pm0.01$ & 0.03$\pm0.01$ & 0.03$\pm0.01$ \\
Strict gap \hfill HERWIG6  & -- & 0.1 \cite{Kepka:2010hu} & -- & 0.1 \cite{Sirunyan:2021oxl} \\
Strict gap \hfill PYTHIA8 & 0.23$\pm$0.09 & 0.15$\pm$0.04 & 0.28$\pm$0.04 & 0.35$\pm$0.09\\
Exp.\ gap CP1 \hfill PYTHIA8 & 0.10$\pm$0.03 & 0.07$\pm$0.02 & 0.08$\pm$0.01 & 0.08$\pm$0.02 \\
Exp.\ gap CP5 \hfill PYTHIA8 & 0.38$\pm$0.12 & 0.25$\pm$0.07 & 0.87$\pm$0.17 & 1.11$\pm$0.22 \\
\hline
KMR Model 4  & 0.028 \cite{Khoze:2013dha}& 0.028 \cite{Khoze:2013dha} & 0.015 \cite{Khoze:2013dha}& 0.010 \cite{Khoze:2013dha} \\
GLM Model I  & 0.0760 \cite{Gotsman:2015aga}& 0.0760 \cite{Gotsman:2015aga}& 0.0363 \cite{Gotsman:2015aga} & 0.0230 \cite{Gotsman:2015aga} \\ 
GLM Model IIn  & 0.0334 \cite{Gotsman:2015aga}& 0.0334 \cite{Gotsman:2015aga}& 0.0310 \cite{Gotsman:2015aga} & 0.0305 \cite{Gotsman:2015aga} \\ 
GLM Model III  & 0.0168 \cite{Gotsman:2015aga}& 0.0168 \cite{Gotsman:2015aga}& 0.0063 \cite{Gotsman:2015aga} & 0.0044 \cite{Gotsman:2015aga} \\ 
\hline
\end{tabular}
\end{table}

The scale factor $\mathcal{S}$ used to normalize our predictions shall not be regarded as a proxy of the survival probability, since it may include in addition the normalization effects from missing higher-order corrections in the calculation. Nevertheless, it is interesting to make a comparison of the $\sqrt{s}$-dependence and size of $\mathcal{S}$ with existing nonperturbative QCD calculations of $\mathcal{S}_\mathrm{prob}$ from diffractive models.


The calculations for survival probabilities $\mathcal{S}_\mathrm{prob}$ are available for single- and central diffractive topologies with low momentum transfers; there are no calculations available for the jet-gap-jet topology with large momentum transfers. The comparison deals only with the $\sqrt{s}$ variation of $\mathcal{S}$ or $\mathcal{S}_\mathrm{prob}$ assuming that the additional factors between $\mathcal{S}$ and $\mathcal{S}_\mathrm{prob}$ do not depend strongly on $\sqrt{s}$ or on kinematics.

The different numbers for the CDF and D0 experiments at the Tevatron with 1.8 TeV center-of-mass energy as well as for the CMS experiment at center-of-mass energies of 7 and 13 TeV are shown in Table~\ref{tab:03}. We stress that the values of $\mathcal{S}$ we quote depend on the rapidity gap definition, and therefore the interpretation must be made with care for each case. 
The first line refers to our full BFKL results, where no gap condition was imposed. Consequently, a smaller suppression factor of about 0.03 is required to make the PYTHIA8 simulations consistent with the data. As mentioned above, there is no obvious energy dependence. With a strict gap definition (i.e.\ no particles allowed in the gap) applied to HERWIG6 simulations (line two), RMK obtained the same scale factor of 0.1 for D0 \cite{Kepka:2010hu} as they did before with their partonic calculation. The same value of $\mathcal{S}$ 
was shown in the experimental CMS paper to describe also the 13 TeV data \cite{Sirunyan:2021oxl}, i.e.\ there was no center-of-mass energy dependence. This is confirmed by our PYTHIA8 simulations, in particular when the experimental gap definition (fourth row) is applied. With multiparton interactions, PYTHIA8 is able to describe the data (within errors) without a static additional suppression factor at all center-of-mass energies (last row).

Several theoretical groups have made predictions for the survival probability at the Tevatron and LHC based on two-channel eikonal models, in particular Khoze--Martin--Ryskin (KMR) ~\cite{Khoze:2013dha} and Gotsman--Levin--Maor (GLM)~\cite{Gotsman:2015aga}. While both theoretical approaches are based on the Good-Walker formalism \cite{Good:1960ba} and conceptually similar, they differ in their approximations and statistical procedures. The Durham group (KMR) takes into account all $n\to m$ pomeron transitions in the framework of a partonic model and sums all diagrams by numerical solution of a system of highly non-linear equations for amplitudes \cite{Khoze:2013dha}. To account for semi-hard and hard interactions, three types of pomeron poles are introduced. Formulae for cross sections of different inelastic diffractive processes are obtained using probabilistic arguments and not cutting rules as in the standard approach. In this model, it is possible to obtain a reasonable description of the total cross section for $pp$ interactions, the elastic cross section in the diffraction cone region, and cross sections of single and double diffraction. A different approach is used by the Tel-Aviv group (GLM) \cite{Gotsman:2015aga}. Arguments, based on a small value of the pomeron slope, are used to justify applicability of pQCD for diffractive processes. Motivated by pQCD, the authors use the triple-pomeron interaction only with the maximal number of pomeron loops. The main ingredient is the BFKL pomeron Green's function, obtained using a color-glass-condensate/saturation approach from the solution of the non-linear Balitsky--Kovchegov equation \cite{Balitsky:1998ya,Kovchegov:1999yj} using the Mueller--Patel--Salam--Iancu approximation \cite{Mueller:1996te,Iancu:2003zr} to sum enhanced diagrams. Also in this model, a good description of the elastic and diffractive cross sections, of inclusive production and the rapidity correlations at high energies can be obtained. We stress, however, that neither model has been designed for or tested against processes with a central rapidity gap and a large momentum transfer.

\begin{figure}
\centering
\includegraphics[width=0.7\textwidth]{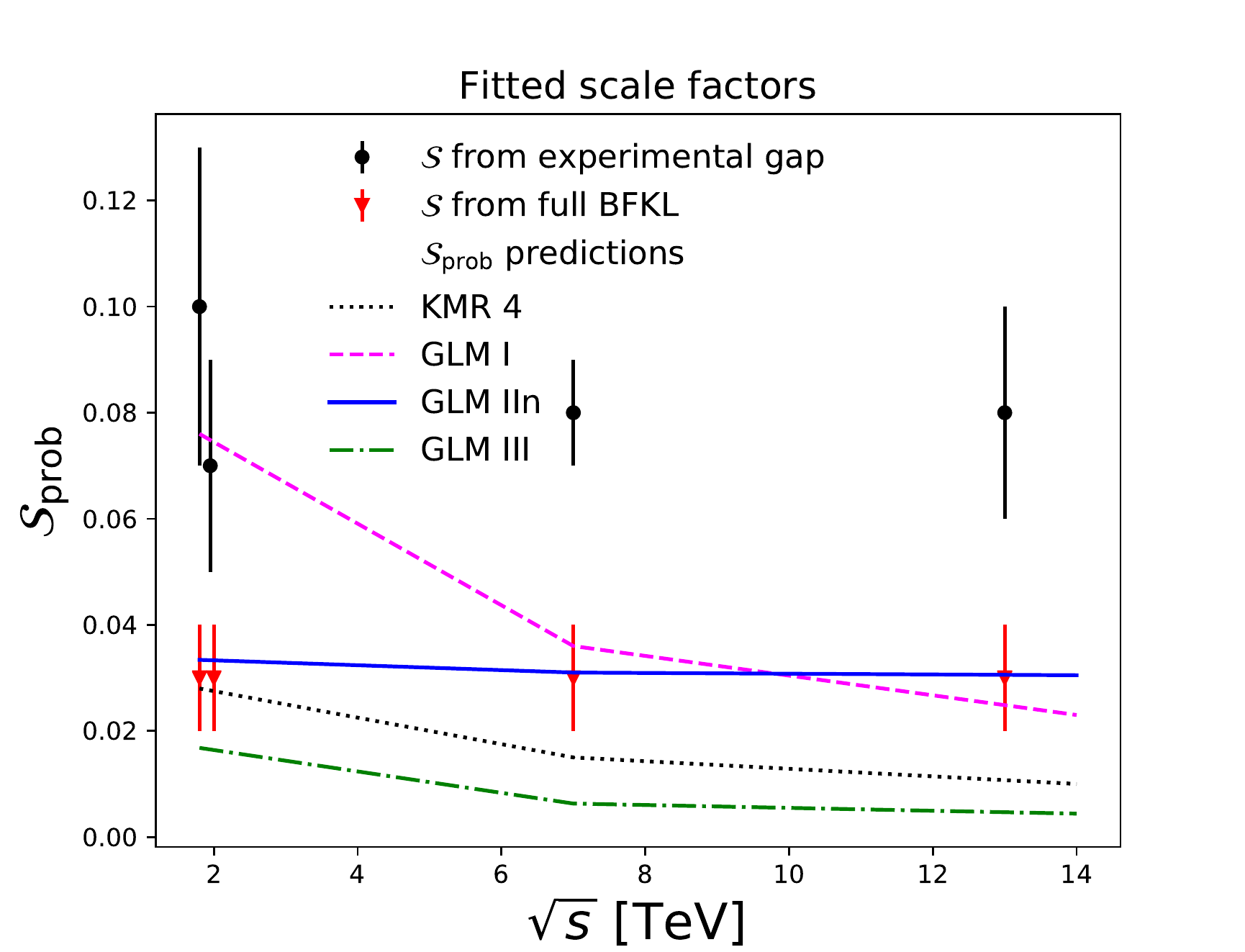}
\caption{Fitted scale factors ${\cal S}$ as a function of center-of-mass energy $\sqrt{s}$ with the CP1 tune of PYTHIA8 without multiparton interactions for the experimental gap definition (black points) and the full BFKL prediction (red points), compared to theoretical predictions for the survival probability $\mathcal{S}_\mathrm{prob}$ by KMR (black line) \cite{Khoze:2013dha} and GLM (other colors) \cite{Gotsman:2015aga}.}
\label{fig:09}
\end{figure}

Using ATLAS inelastic cross section data at 7 TeV as a function of rapidity and elastic data from 62.5 GeV to 7 TeV, KMR obtain four fits for the survival probability. Since their Model 4 also describes the TOTEM data at 7 TeV very well, it is favored by the authors and thus listed in Table~\ \ref{tab:03}. It is characterized by small values which only agree marginally with our full BFKL prediction (red points). Furthermore, the KMR 4 predictions fall off with center-of-mass energy as opposed to our findings. This can also be seen in Fig.\ \ref{fig:09} (black dotted line). GLM fits a large body of LHC data with center-of-mass energies up to 7 TeV and proposes three models based on different approximations. They also updated the parameters of one of the models, namely Model IIn. Their Model III (dot-dashed green line) is closest to the KMR 4 prediction. Model I (dashed purple line) has the largest values at 1.8 TeV and is thus the model closest to our suppression factor for an experimental gap definition (black points), but at the same it shows the strongest decrease with center-of-mass energy. So overall the best agreement is found between our full BFKL results (red points) and the updated GLM Model IIn (full blue line), which both have little observed energy dependence. Note that both the KLM and GLM predictions for the highest LHC energy are actually for 14 TeV, not 13 TeV.

\section{Summary and outlook}
\label{sec:6}

In summary, we have presented in this paper a phenomenological analysis of events with two high transverse momentum ($p_T$) jets separated by a large (pseudo-)rapidity interval void of particle activity, also known as jet-gap-jet events. We embedded the BFKL pomeron exchange amplitudes, with resummation at the next-to-leading logarithmic (NLL) approximation, in the PYTHIA8 Monte Carlo event generator and simulated standard QCD dijet events next-to-leading order (NLO) in $\alpha_s$ matched to parton showers with POWHEG+PYTHIA8. We compared our calculations to measurements by the CDF, D0 and CMS experiments at center-of-mass energies of 1.8, 7 and 13~TeV, putting special emphasis on the impact of the theoretical scales, the parton densities, final- and initial-state radiation effects, multiple parton interactions, and $p_T$ thresholds and multiplicities of the particles in the rapidity gap. We found that with a strict gap definition, the shapes of most distributions are well described except for the CMS azimuthal-angle distribution at 13 TeV, which was best described by the full BFKL prediction. The survival probability was surprisingly well modelled by multiparton interactions in PYTHIA8. Without multiparton interactions, theoretical predictions for central-diffractive and single-diffractive topologies based on two-channel eikonal models agreed qualitatively with fits to the experimental data.

Initial-state radiation plays a significant role in the definition of the rapidity gap between the jets. It produces numerous colored charges in the forward and backward pseudorapidities, which are then color reconnected with the colored charges produced in the forward and backward regions. When these long color strings are broken, they lead to the formation of several hadrons in the gap region between the final-state jets. In other words, there is net color flow in this case, even in the presence of hard $t$-channel CSE. It is likely that the effect we are seeing is mostly due to the fragmentation model used in PYTHIA8, as well as the way the ISR parton showers are described for quark- and gluon-initiated processes. For future studies, one could test these effects by implementing the Mueller--Tang amplitudes in other general purpose MC generators with different fragmentation models and different ways of implementing the ISR parton shower, for example in the latest versions of the HERWIG \cite{Bellm:2019zci} or SHERPA \cite{Sherpa:2019gpd} event generators.

Note that the CMS tunes used for our predictions \cite{CMS:2019csb} have been validated with 13-TeV charged-particle spectra of minimum-bias inelastic, non-single diffractive (NSD) and single-diffractive (SD) events \cite{CMS:2018nhd} sensitive to SD, CD, and DD dissociation, but not with central-gap topologies. We therefore propose that for future measurements for Monte Carlo tuning, topologies with such central gaps should be included in order to properly tune ISR. This could also affect other topologies with central gaps that are used at high-energy colliders (e.g., vector boson fusion topologies). It would even be interesting to simply measure the charged-particle spectra for minimum-bias events when applying such a strict gap condition as we do for jet-gap-jet events and compare them to PYTHIA8 with different tunes. It is well possible that the existing tunes turn out to be inadequate for this topology.

For future phenomenological investigations, one could take into account the Mueller--Tang NLO impact factors in the theoretical calculation. Progress in this direction has been presented in Refs.~\cite{hentschinski1,hentschinski2, Deganutti:2017usx, Deganutti:2020zzf}. These corrections might modify, in a non-trivial way, the dependence of the CSE cross section with the jet kinematics. Another ingredient that is missing in the phenomenological calculation is the effect of wide-angle, soft-gluon emissions into the rapidity gap region between the jets. For final-states with energy flow veto, these soft-gluon emissions lead to non-global logarithms that need to be resummed at all orders in the perturbative expansion~\cite{uedahatta}. The resummation of these non-global logarithms is absent in the BFKL framework, but prescriptions to implement them in the Mueller--Tang jets phenomenology have been presented in Ref.~\cite{uedahatta}.

Finally, to complement the jet-gap-jet process in proton-(anti-)proton collisions, it would be interesting to use other probes with central rapidity gaps that have different sensitivities to initial-state radiation and soft particle activity. For example, $\mathrm{J}/\psi$-gap-$\mathrm{J}/\psi$ events might be an interesting venue for such investigations (or other quarkonia pairs). There are no final-state radiation parton showers, and such a process is expected to be gluon initiated at the lowest order in perturbation theory. The $\mathrm{J}/\psi$-gap-$\mathrm{J}/\psi$ process could be investigated in ultraperipheral heavy-ion collisions as well~\cite{Kwiecinski:1998sa, Gon_alves_2006}, completely removing the dependence on the modeling of initial-state radiation, color reconnections, underlying event activity, and final-state radiation. The measurement of the jet-gap-jet process in proton-proton collisions with forward intact protons is also interesting, since the soft parton exchanges are suppressed due to the intact proton(s) condition. The CMS and TOTEM experiments demonstrated that the $f_\text{CSE}$ fraction is larger in events with at least one intact proton~\cite{Sirunyan:2021oxl}, but with the limitation of not having the possibility to perform a differential analysis. In principle, one could use these events to study BFKL dynamics in an environment with a strong suppression of MPI.

\acknowledgments
We thank F.\ Deganutti, A.\ Ekstedt, R.\ Enberg, G.\ Ingelman, L.\ Motyka, and P.\ Risse for helpful discussions. This work has been supported by the DFG through the Research Training Network 2149 ``Strong and weak interactions - from hadrons to dark matter'' and through Project-Id 273811115 - SFB 1225 ``ISOQUANT'' as well as by the Alexander von Humboldt foundation through a Research Award. The calculations have been performed on the high-performance computing cluster PALMA II at WWU Münster.

\appendix
\section{The S3 and S4 schemes for non-zero conformal spins}
\label{app:a}

The effective characteristic function $\chi_{eff}$ entering the BFKL gluon scattering
amplitude in Eq.\ (\ref{jgjnll}) at LL order is
\bea
\chi_{LL}(p,\gamma)&=&2\psi(1)-\psi(1-\gamma+\frac{|p|}{2})-\psi(\gamma+\frac{|p|}{2})
\label{chill}
\eea
with $\bar\alpha=\alpha_s N_c/\pi=$ const.\ and $\psi(x)=\Gamma'(x)/\Gamma(x)$ the logarithmic derivative of the Euler gamma function $\Gamma(x)$.

Beyond LL, a regularized result can be obtained from the one at NLL by solving the implicit equation \cite{Salam:1998tj,Vera:2007kn,Schwennsen:2007hs}
\bea
\chi_{eff}&=&\chi_{\rm NLL}(p,\gamma,\bar\alpha\ \chi_{eff}).
\label{eff}
\eea
The scale-invariant and $\gamma \leftrightarrow 1-\gamma$ symmetric part of the NLL function
extended to non-zero conformal spins $p\neq0$ is
\bea
\chi_1(p,\gamma)&=&
\frac{3}{2}\zeta(3)
+\lr \frac{1+5b}{3}-\frac{\zeta(2)}2\rr\chi_{LL}(p,\gamma)
-\frac{b}{2}\chi_{LL}^2(p,\gamma)\nonumber\\
&+&\frac{1}{4}\left[\psi''\lr \gamma+\frac{p}{2}\rr+\psi''\lr 1-\gamma+\frac{p}{2}\rr\right]
-\frac{1}{2}\left[\phi(p,\gamma)+\phi(p,1-\gamma)\right]\nonumber\\
&-&\frac{\pi^2\cos(\pi\gamma)}{4\sin^2(\pi\gamma)(1-2\gamma)}
\left\{\left[3+\lr 1+\frac{N_f}{N_c^3}\rr\frac{2+3\gamma(1-\gamma)}{(3-2\gamma)(1+2\gamma)}\right]\delta_{0p}
\right.\nonumber\\
&& \hspace*{40mm}\left.-\lr 1+\frac{N_f}{N_c^3}\rr\frac{\gamma(1-\gamma)}{2(3-2\gamma)(1+2\gamma)}\delta_{2p}\right\},
\eea
with $\bar\alpha(k^2)=\alpha_s(k^2)N_c/\pi=\left[b\log(k^2/\Lambda_{QCD}^2)\right]^{-1}$,
$b=(11N_c-2N_f)/(12N_c)$, and
\bea
\phi(p,\gamma)=\sum_{k=0}^\infty\frac{(-1)^k}{k+\gamma+p/2}\left\{
\psi'(k+1)-\psi'(k+p+1)+\frac{\psi(k+p+1)-\psi(k+1)}{k+\gamma+p/2}\right.\nonumber\\\left.
+\frac{(-1)^k}4\left[\psi'\lr \frac{k+p+2}{2}\rr
-\psi'\lr \frac{k+p+1}{2}\rr+\psi'\lr \frac{k+2}{2}\rr-\psi'\lr \frac{k+1}{2}\rr\right]\right\}.
\label{klcor}
\eea
Note that for the terms on the first line of Eq.~\eqref{klcor} inside the curly brackets, we have corrected the signs with respect to Ref.\ \cite{Kotikov:2000pm}, where they are misprinted (the signs are correct in Ref. \cite{Kotikov:2002ab}). As is the case for 
$\chi_{LL}(p,\gamma),$ the kernel $\chi_1(p,\gamma)$ has poles at $\gamma=-p/2$ and $\gamma=1+p/2.$ The pole structure at $\gamma=-p/2$ (and by symmetry at $\gamma=1+p/2$) is
\beq
\chi_1(p,\gamma)=-\frac{1}{2\lr \gamma+\frac{p}{2}\rr^3}+\frac{d_2(p)}{\lr \gamma+\frac{p}2\rr^2}
+\frac{d_1(p)}{\lr \gamma+\frac{p}2\rr}+{\cal O}(1)
\eeq
with
\bea
d_1(p)&=&\frac{1+5b}3-\frac{\pi^2}8+b[\psi(p+1)-\psi(1)]+
\frac{1}{8}\left[\psi'\lr \frac{p+1}2\rr-\psi'\lr \frac{p+2}2\rr+4\psi'\lr p+1\rr\right]
\nonumber\\
&-&\lr 67+13\frac{N_f}{N_c^3}\rr\frac{\delta_{0p}}{36}
-\lr 1+\frac{N_f}{N_c^3}\rr\frac{47\delta_{2p}}{1800}
\eea
and
\beq
d_2(p)=-\frac{b}2-\frac{1}{2}[\psi(p+1)-\psi(1)]
-\lr 11+2\frac{N_f}{N_c^3}\rr\frac{\delta_{0p}}{12}-\lr 1+\frac{N_f}{N_c^3}\rr\frac{\delta_{2p}}{60}\ .
\eeq
Note that $\chi_1(2,\gamma)$ also has a pole at $\gamma=0$ with residue $(1+N_f/N_c^3)/24.$ This manifestation of the non-analyticity \cite{Kotikov:2000pm} of $\chi_1(p,\gamma)$ with respect to conformal spins does not alter the stability of the NLL prediction and a careful treatment of this singularity is not required.

In the regularization procedure of Ref.\ \cite{Salam:1998tj}, the freedom in the choice of the divergent functions results in differences in the kernel at higher orders. In the S3 scheme, the kernel up to NLL order is given by 
\bea
\chi_{S3}
&=&[1\!-\!\bar\alpha A(p)]
\left[2\psi(1)
\!-\!\psi\lr \gamma+\frac{p+2\bar\alpha B(p)+\omega}2\rr
\!-\!\psi\lr 1-\gamma+\frac{p+2\bar\alpha B(p)+\omega}2\rr\right]
\\ &+&\bar\alpha\left\{\chi_1(p,\gamma)+A(p)\chi_{LL}(p,\gamma)
+\lr B(p)+\frac{\chi_{LL}(p,\gamma)}2\rr\left[\psi'\lr \gamma+\frac{p}2\rr+\psi'\lr 1-\gamma+\frac{p}2\rr\right]\right\} \nonumber
\eea
with $A(p)$ and $B(p)$ chosen to cancel the singularities of $\chi_1(p,\gamma)$ at 
$\gamma=-p/2:$
\beq
A(p)=-d_1(p)-\psi'(p+1)\ ,\hspace{1cm}B(p)=-d_2(p)+\frac{1}{2}[\psi(p+1)-\psi(1)]\ .
\eeq
In contrast, in the S4 scheme the kernel is given by 
\bea
\chi_{S4}
&=&\chi_{LL}(p,\gamma)-f(p,\gamma)
+[1-\bar\alpha A(p)]f(p+\omega+2\bar\alpha B(p),\gamma)
+\bar\alpha\Big\{\chi_1(p,\gamma)
\nonumber\\&+& A(p)f(p,\gamma)
+\lr B(p)+\frac{\chi_{LL}(p,\gamma)}2\rr
\left[\lr \gamma+\frac{p}2\rr^{-2}+\lr 1-\gamma+\frac{p}2\rr^{-2}\right]\Big\}
\eea
with
\beq
f(p,\gamma)=\frac{1}{\gamma+\frac{p}2}+\frac{1}{1-\gamma+\frac{p}2}\ .
\eeq
In this scheme, $A(p)$ and $B(p)$ are given by:
\beq
A(p)=-d_1(p)-\frac{1}{2}\left[\psi'(p+1)-\psi'(1)+\frac{1}{(p+1)^2}\right]
 \ , \ B(p)=-d_2(p)+\frac{1}{2}[\psi(p+1)-\psi(1)]\ .
\eeq

\begin{figure}
\centering
\includegraphics[width=0.8\linewidth]{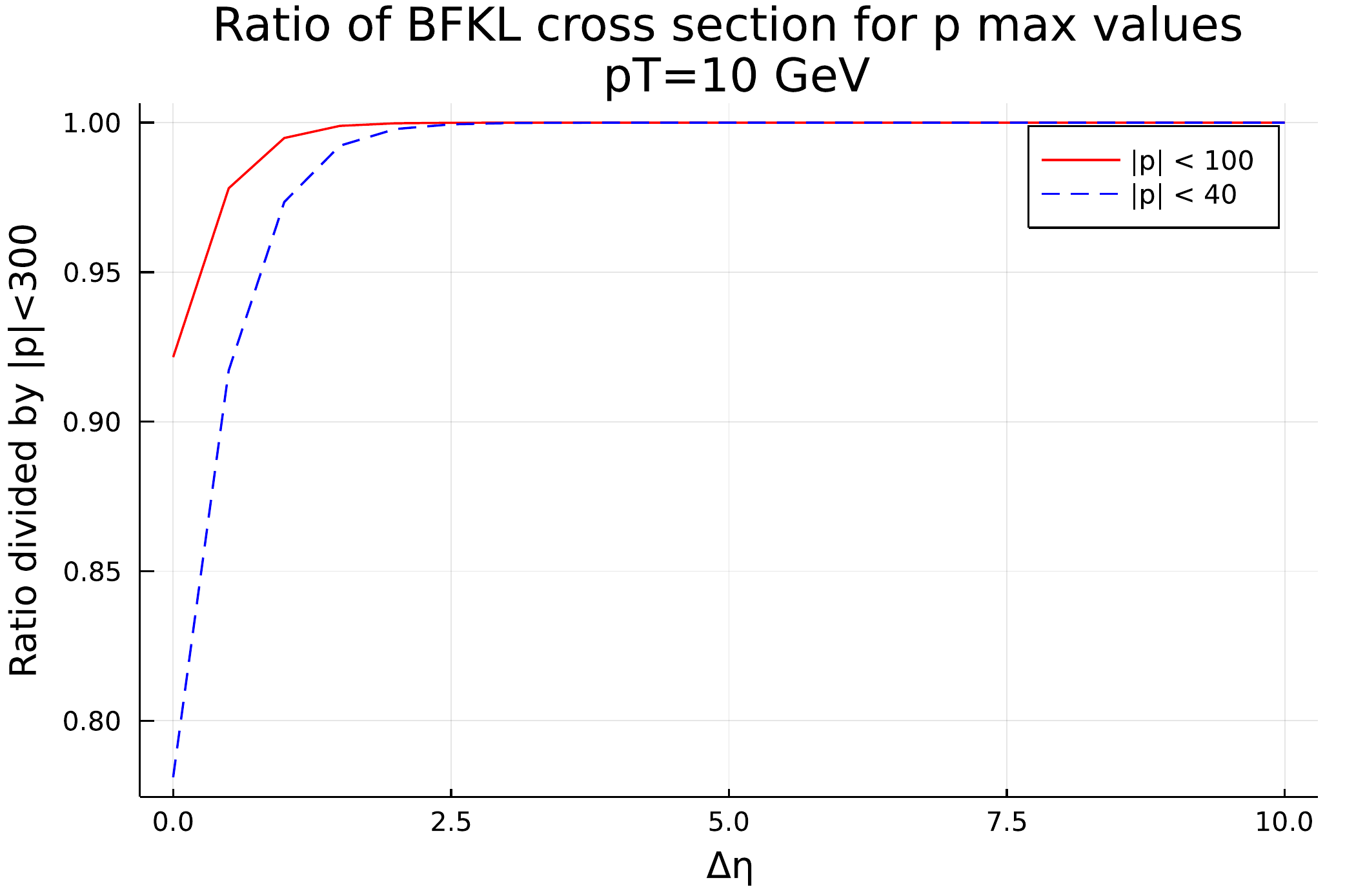}
\caption{NLL BFKL cross section ratios summing up contributions up to $|p|<40$ (blue dashed line) and $|p|<100$ (orange full line) with respect to the one summed to $|p|<300$. The contribution from $|p|=300$ is about four orders of magnitude smaller than the one from $|p|=0$.}
\label{fig:10}
\end{figure}

Convergence of the conformal spin series requires more terms for lower values of $\Delta\eta$ ($\Delta \eta = \Delta y$ at parton-level). This can be seen from Fig.\ \ref{fig:10}, where we plot the ratios of NLL BFKL cross sections in the S4 scheme summing up contributions up to $|p|<40$ (blue dashed line) and $|p|<100$ (orange full line) with respect to the one summed to $|p|<300$. There the contributions are about four orders of magnitude smaller than the one from $|p|=0$. Very small values of $\Delta\eta$ are, of course, outside the domain of validity of the BFKL calculation. Since the NLO impact factors benefit from adding up fewer terms, we use $|p|<40$ as a compromise.

\begin{figure}
\centering
\includegraphics[width=0.8\linewidth]{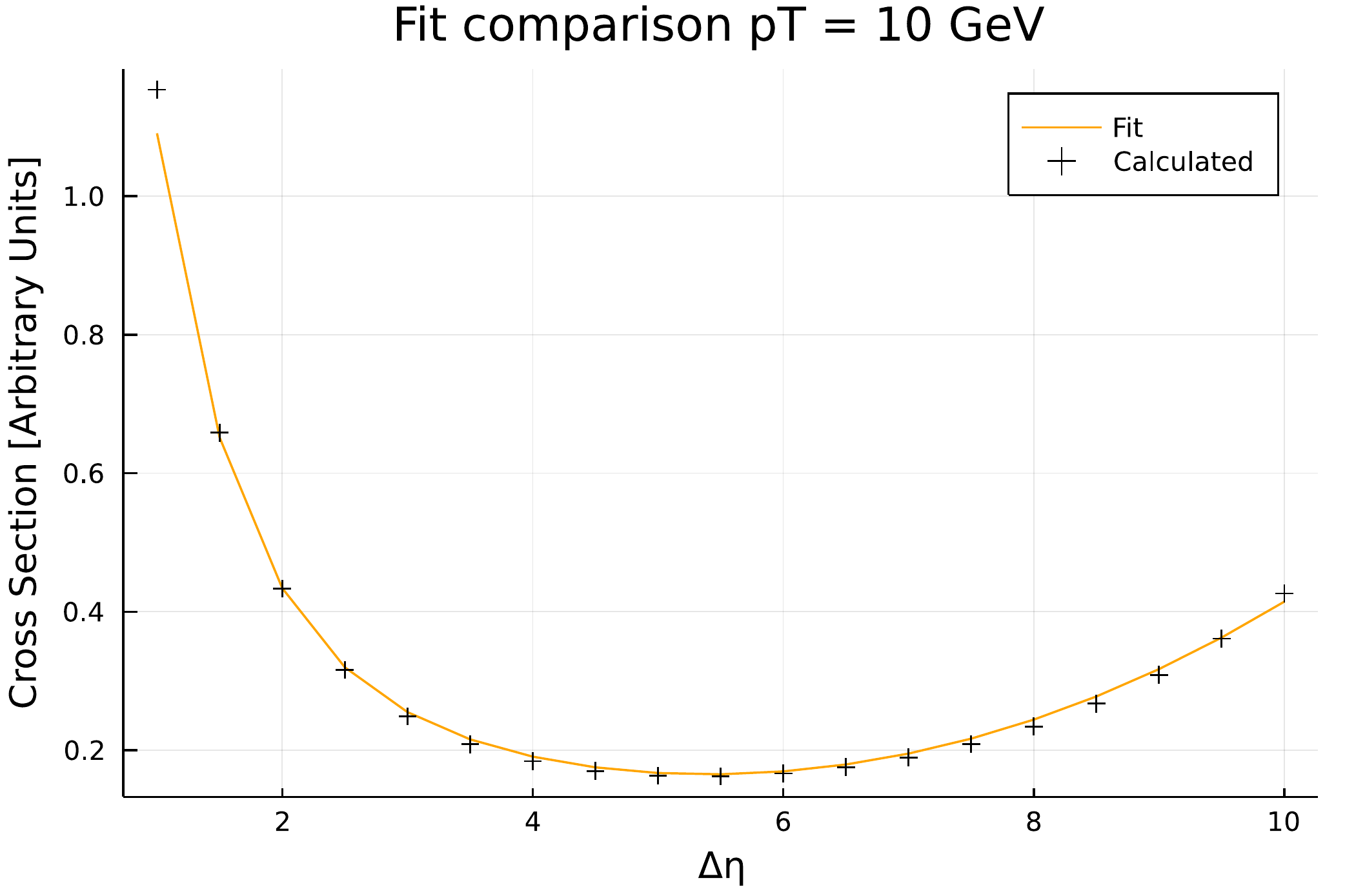}
\caption{NLL BFKL cross section as a function of $\Deta$ calculated with Eq.\ \eqref{jgjnll} (crosses) and our fit (orange line) with the parametrization in Eq.\ (\ref{formulafit}).}
\label{fig:11}
\end{figure}

Since the sum over $|p|$ and the integral over $\gamma$ in Eq.\ (\ref{jgjnll}) are too time-consuming for MC simulations with PYTHIA8, we parametrize the differential parton-level cross section $d\sigma/dp_T^2$ as a function of the parton $p_T$ and $\Delta\eta$ between both partons at generator level. Denoting $z(p_T^2)=\bar\alpha(p_T^2)\Delta\eta/2$, the (purely phenomenological) parametrization used is
\bea
\frac{d \sigma}{dp_T^2}&=&\frac{\alpha_S^4(p_T^2)}{4\pi p_T^4}  \left[ a + b p_T + c \sqrt{p_T}
+ (d + e p_T + f \sqrt{p_T})\times z + (g + h p_T)\times z^2 \right. \nonumber\\ &+& \left.
(i + j \sqrt{p_T})\times z^3 + \exp(k + l z) \right]\ .
\label{formulafit}
\eea
It is then fitted to the full expression of $d\sigma/dp_T^2$ in the S4 scheme in the ranges $p_T\in[10;120]$~GeV and $\Delta\eta \in [1;10]$. The result for $p_T=10$ GeV is shown in Fig.\ \ref{fig:11}.

\begin{table}[t]
\caption{Parameter values using formula~\ref{formulafit} to fit the Next-to-Leading-Log BFKL cross section obtained with the S4 scheme.}
\label{tab:appA1}
\centering
\begin{tabular}{ *{6}{|c}| } 
    \hline
    a & b & c & d & e & f\\
    47.414 & 0.0072066 & 1.5660 & -121.50 & -0.29812 & -3.1149 \\
    \hline\hline
     g & h & i & j & k & l \\
     119.93 & 0.55726 & 10.385 & 1.3812 & 5.9833 & -17.199 \\
    \hline
\end{tabular}
\end{table}

%

\section{Inclusive dijet cross section at 13 TeV}
\label{app:b}

\begin{figure}
\centering
\includegraphics[width=0.496\linewidth]{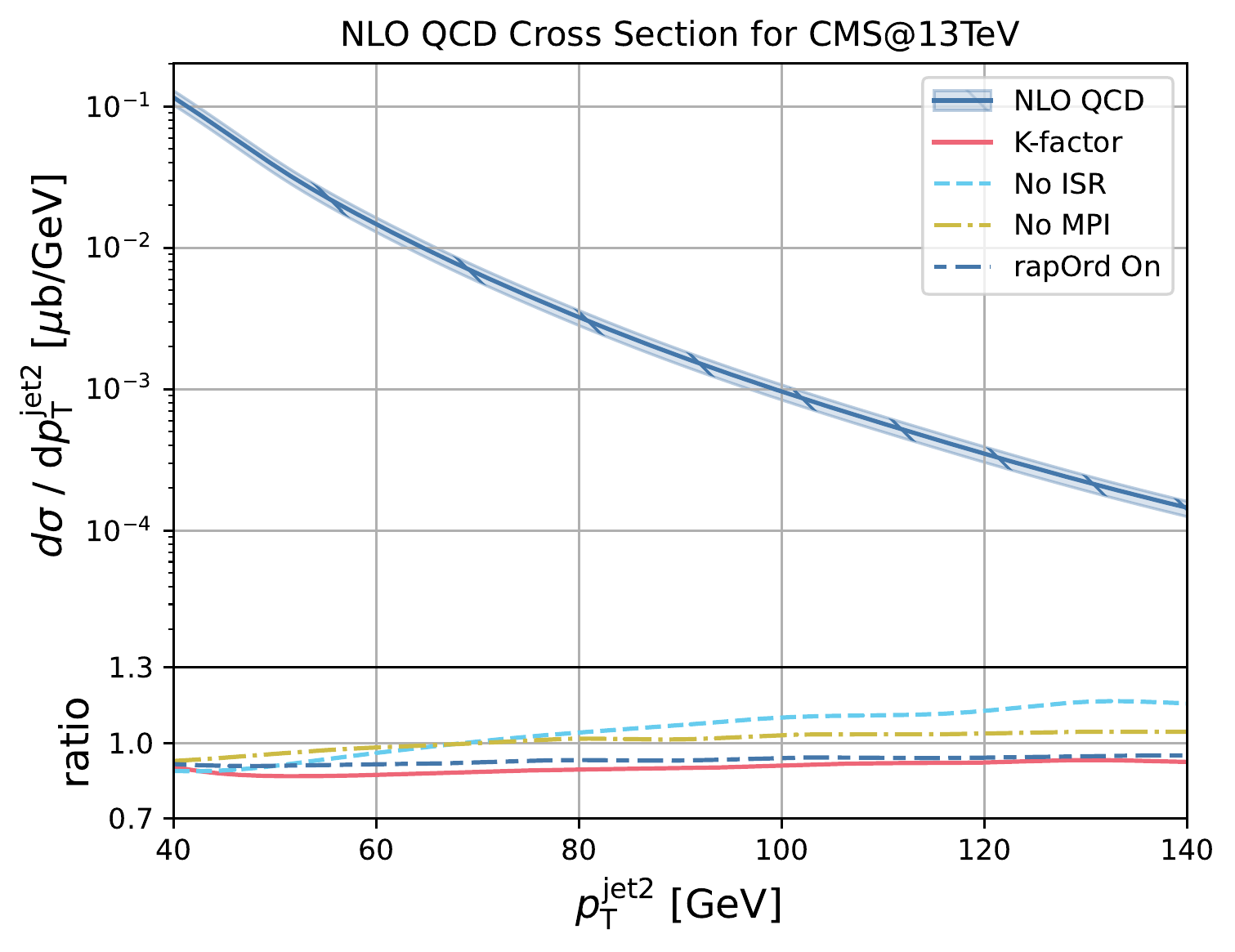}
\includegraphics[width=0.496\linewidth]{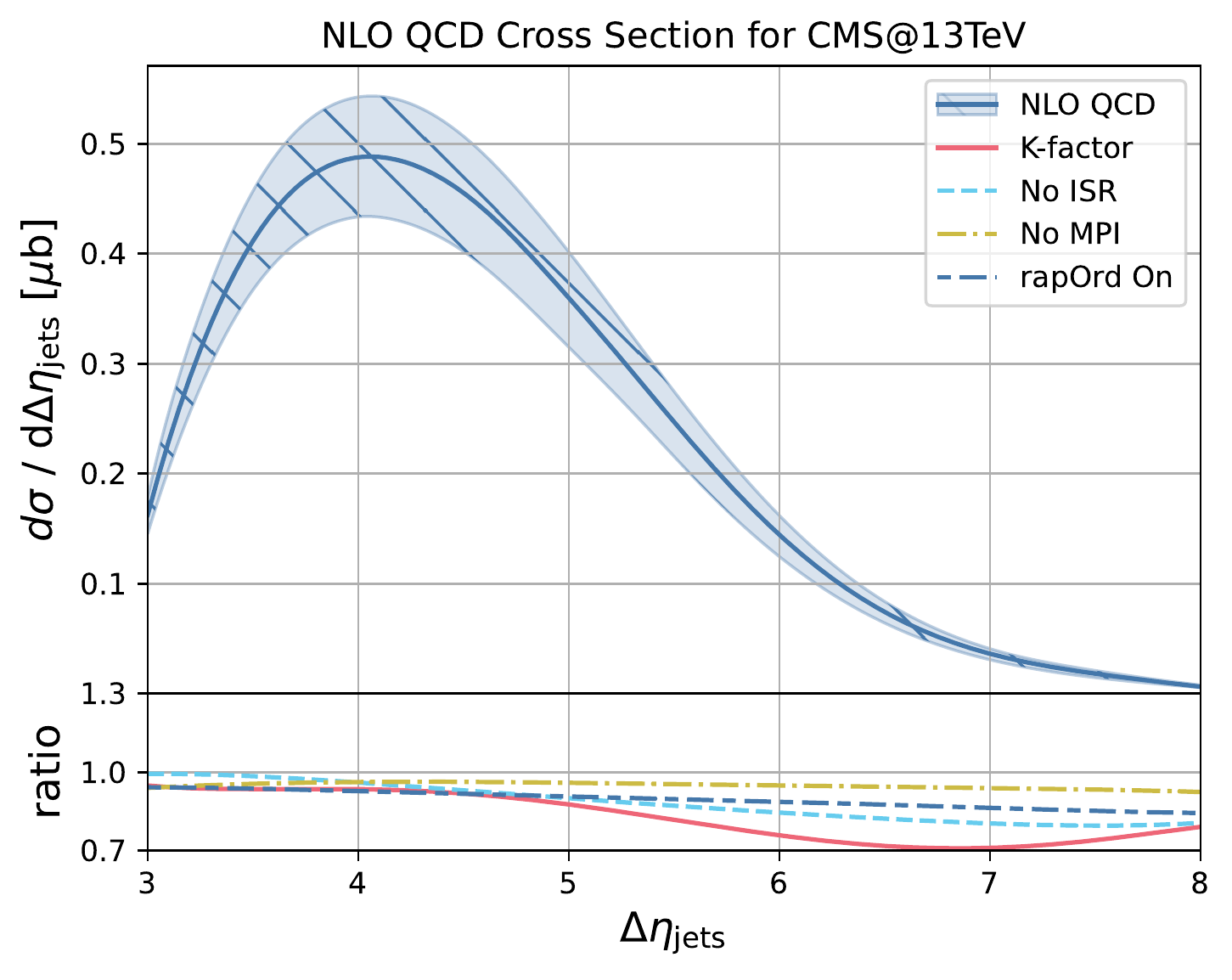}
\includegraphics[width=0.515\linewidth]{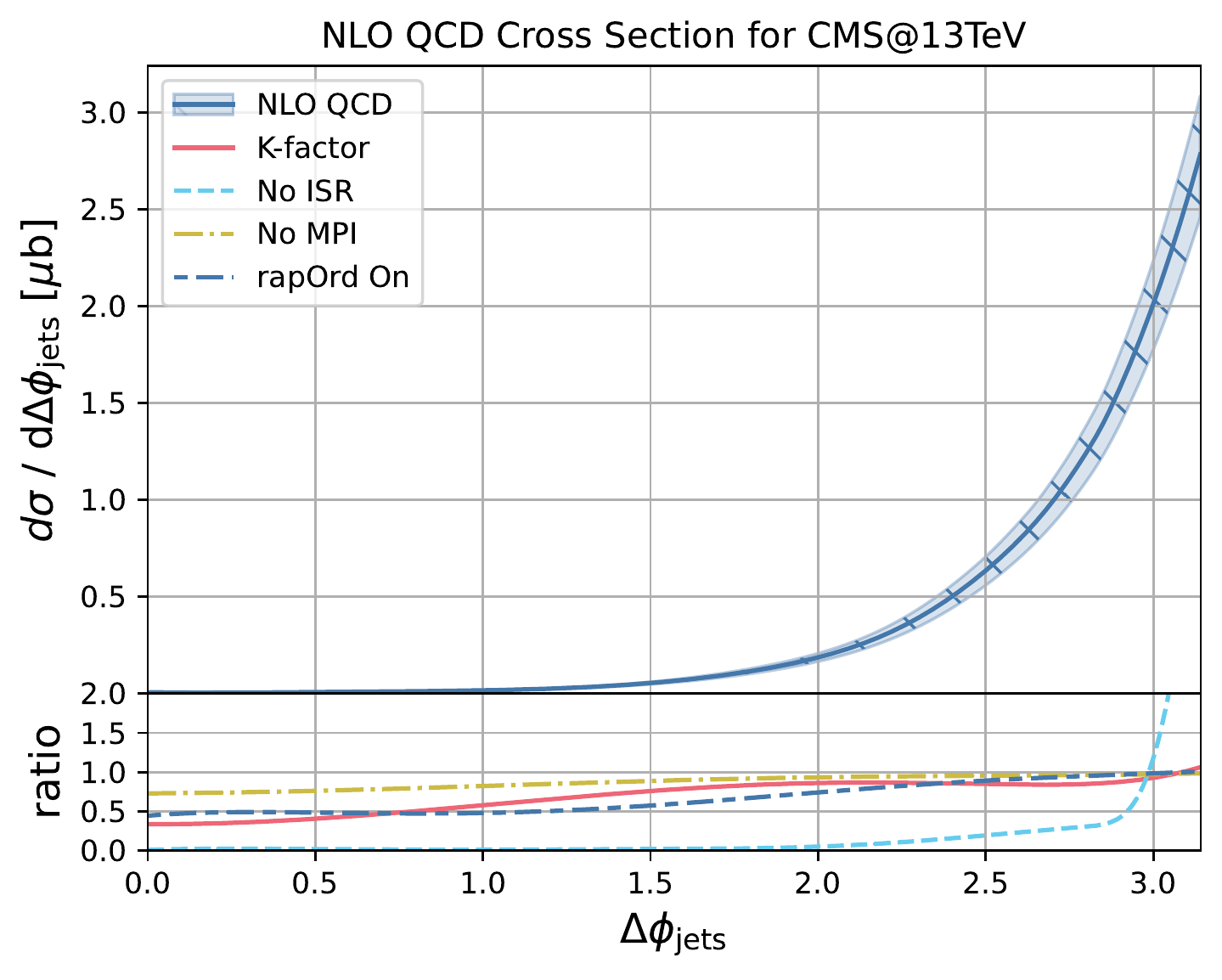}
\caption{Inclusive dijet cross sections for CMS conditions at a center-of-mass energy of 13 TeV as a function of $\pTj$, $\Deta$, and $\Dphi$. The upper panels show the absolute cross section at NLO+PS and their scale uncertainty based on the seven-point method. The lower panels show the NLO/LO $K$-factor (red) as well as the impact of initial-state radiation (ISR, light blue), multi-parton interactions (MPI, yellow) and rapidity ordering in the parton shower (rapOrd On, dark blue) in LO PYTHIA8 simulations as a ratio with respect to using the CP1 tune.}
\label{fig:03}
\end{figure}

For future reference, we present in this appendix and Fig.\ \ref{fig:03} the inclusive dijet cross section at the LHC with a center-of-mass energy of $\sqrt{s}=13$ TeV in the kinematics of the CMS jet-gap-jet analysis \cite{Sirunyan:2021oxl}. The corresponding experimental cuts are listed in Tab.\ \ref{tab:02}. The predictions have been obtained at NLO matched to parton showers and hadronization with POWHEG \cite{Alioli:2010xa} and PYTHIA8 \cite{Sjostrand:2014zea}, using the CMS NLO tune CP3 (see Tab.\ \ref{tab:01}) \cite{CMS:2019csb} and renormalization and factorization scales set to the jet transverse momentum $p_T$ (upper panels, full blue curves). The scale uncertainties (shaded blue bands) are obtained with the seven-point method, i.e.\ by varying the scales independently by relative factors of two, but not four around the central scale.

The distribution in the transverse momentum of the second-hardest jet $p_\mathrm{T}^\mathrm{jet2}$ (top left) starts at the cut (40 GeV) and falls off with $1/(p_\mathrm{T}^\mathrm{jet2})^4$ as expected. The distribution in pseudorapidity difference $\Delta \eta_{\rm jets}$ (top right), constrained to lie within 2.8 and 9.4, peaks around four, i.e.\ close to the minimum allowed by the cuts, due to the strong connections from color-octet exchanges. The distribution in relative azimuthal angle $\Delta\phi_{\rm jets}$ (bottom) is sharply peaked at $\pi$ as expected for back-to-back jets in the LO configuration, which is smeared out by initial-state radiation at NLO and from the PS (see below).

The lower panels show the impact of various higher-order QCD contributions. In particular, we show the $K$-factor of NLO over LO predictions (full red curves), where the latter still include PYTHIA8 parton showers and hadronization and have been obtained with the CMS tune CP1 (see Tab.\ \ref{tab:01}). In general, the $K$-factor of about 0.9 is slightly smaller than one, as expected for an analysis with a small jet radius $R$ (here 0.4). The impact of the NLO corrections decreases slightly with increasing transverse momentum (top left) and renormalization scale, since the strong coupling is then smaller. The increase of the NLO effects with the pseudorapidity difference (top right) is more significant, since the hierarchy of two relevant scales $p_T$ and dijet invariant mass $M=2p_T\cosh(\Delta\eta_{\rm jets}/2)$ becomes more pronounced. The increase of the NLO effects towards small $\Delta\phi_{\rm jets}<\pi$ (bottom) is even larger, as the LO distribution has support only in the back-to-back configuration.

The full NLO+PS $K$-factor can be compared to the impact of the PS alone, which alters only the differential shape, but not the total normalization of the cross sections. For consistency, these comparisons are performed at LO, since at NLO the PS is matched to the NLO calculation in POWHEG. Final-state radiation (FSR, not shown) changes neither the $p_\mathrm{T}^\mathrm{jet2}$- nor the $\Delta\eta_{\rm jets}$-distributions significantly, since the radiation gets recombined into the jet, but it does broaden the $\Delta\phi_{\rm jets}$-distribution, when it happens out of the two (here relatively narrow) jet clusters. The overall normalization decreases, as the FSR outside the cluster effectively shifts the jet energy scales for quark and gluon jets. In the small-$R$ approximation \cite{Aversa:1988vb,deFlorian:2007fv,Dasgupta:2007wa,CMS:2020caw},
\bea
 (\delta p_T)_q &=& -{\alpha_s p_T\over\pi}\ln\lr{1\over R}\rr\le C_F\lr2\ln2-{3\over8}\rr\re+{\cal O}(\alpha_s)\sim -2.5~{\rm GeV},\nonumber\\
 (\delta p_T)_g &=& -{\alpha_s p_T\over\pi}\ln\lr{1\over R}\rr\le C_A \lr2\ln2-{43\over96}\rr+T_Rn_f{7\over48}\re+{\cal O}(\alpha_s)\sim -7~{\rm GeV},\nonumber
\eea
with $C_A=3$, $C_F=4/3$, $T_f=1/2$, and $n_f=5$, so that the distributions are shifted to lower $p_T$, thus reducing the cross section after applying the $p_T$-cut.

In contrast, initial-state radiation (ISR, dashed light-blue curves) only suppresses the production of low-$p_T$ jets (top left) and substantially decreases the fraction of jets separated by a large rapidity gap (top right). The effect is smaller when the ISR is forced to be ordered in rapidity (wide-dashed dark-blue lines). Soft multiparton interactions (MPI, dot-dashed yellow lines) have an even smaller effect as expected. Without ISR, the jets are essentially back-to-back in $\Delta\phi_{\rm jets}$ as enforced by the LO $2\to2$ scattering process (bottom). 

\bibliographystyle{JHEP}
\bibliography{references}

\end{document}